\documentclass[
    prl, twocolumn, superscriptaddress, nofootinbib, amsmath, amssymb,
    aps, floatfix, preprintnumbers
]{revtex4-2}

\usepackage[utf8]{inputenc}
\usepackage{setspace}
\usepackage{graphicx}
\usepackage{booktabs}
\usepackage[dvipsnames]{xcolor}
\usepackage{tikz}
\definecolor{CiteBlue}{RGB}{45,52,151}
\usepackage[
    colorlinks=true,
    linkcolor=CiteBlue,
    urlcolor=CiteBlue,
    citecolor=CiteBlue
]{hyperref}
\usepackage[normalem]{ulem}
\usepackage{cancel} 

\usepackage{physics}
\usepackage{slashed}
\usepackage{graphicx}
\usepackage{dcolumn}
\usepackage{bm}
\usepackage{xcolor}

\input{radial-diagrams.tex}

\begin{document}

\title{Bring Your Own Instanton: \\ Monopole Catalysis from Sourced SU(2) Instantons}

\author{Csaba Cs\'aki}
\email{csaki@cornell.edu}
\affiliation{Department of Physics, LEPP, Cornell University, Ithaca, NY 14853, USA}

\author{Larissa Kiriliuk}
\email{lk559@cornell.edu}
\affiliation{Department of Physics, LEPP, Cornell University, Ithaca, NY 14853, USA}

\author{Zhihan Liu}
\email{zl642@cornell.edu}
\affiliation{Department of Physics, LEPP, Cornell University, Ithaca, NY 14853, USA}

\author{Rotem Ovadia}
\email{rotem.ovadia@mail.huji.ac.il}
\affiliation{Racah Institute of Physics, Hebrew University of Jerusalem, Jerusalem 91904, Israel}

\date{\today}

\begin{abstract}
	We show that Rubakov--Callan catalysis of proton decay in the field of a 't Hooft--Polyakov monopole proceeds through ``sourced $SU(2)$ instantons'', which are novel topological gauge configurations sourced by fermions and carrying unit instanton number.
	Unlike vacuum BPST instantons, these sourced instantons do not give rise to  exponentially suppressed rates.
	We relate the winding number to the number of fermionic zero modes, suggesting an underlying index theorem for sourced instantons.
	We also calculate the closed-form expression for the leading-order correlator for fermion insertions both outside and inside the monopole core.
	Our results indicate that Rubakov--Callan catalysis is an IR effect where the scattering fermions do not reach the monopole core.
\end{abstract}

\maketitle

%%%%%%%%%%%
{\bf Introduction.}---%
%%%%%%%%%%%
Magnetic monopoles are a well-motivated extension of the Standard Model of particle physics.
They can explain electric charge quantization~\cite{Dirac:1931kp} and arise in various theoretical frameworks, including grand unified theories (GUTs)~\cite{tHooft:1974kcl,Polyakov:1974ek}, supersymmetric gauge theories~\cite{Seiberg:1994rs}, and theories with extra dimensions~\cite{Gross:1983hb,Sorkin:1983ns}.
Their existence is also motivated by the magnetic weak gravity conjecture~\cite{Arkani-Hamed:2006emk}.
One of the most striking implications of GUT monopoles is that they can catalyze proton decay through the Rubakov--Callan process~\cite{Rubakov:1981rg,Callan:1982ah,Callan:1982ac} $u^1+u^2+M\to e^\dagger+(d^3)^\dagger+M$, where the superscripts label quark colors.
Proton decay remains one of the primary accessible phenomenological signatures of GUTs.
Despite decades of dedicated searches, no magnetic monopoles have been conclusively observed~\cite{Mitsou:2026zcf}.
Existing bounds nevertheless allow a sufficiently small population of GUT monopoles~\cite{MACRO:2002iaq,Super-Kamiokande:2012tld,IceCube:2014xnp}.

On the theoretical side, monopoles offer a rare opportunity to study non-perturbative dynamics in a controlled setting, since an Abelian gauge sector governs their long-distance interactions.
A longstanding question concerns the interpretation of monopole--fermion scattering in the massless fermion limit, including the proposed process $\bar{e} + M \to (\bar{u}^1)^\dagger + (\bar{u}^2)^\dagger + d^3 + M$.
The question originally arose as a contradiction, sometimes called the unitarity puzzle, in the same two-dimensional reduction that correctly describes the Rubakov--Callan process: the outgoing states appear to carry fractional fermion numbers, giving rise to semitons~\cite{Callan:1982ac}.
Recent approaches include effective boundary conditions at the monopole core~\cite{Brennan:2021ewu,Bolognesi:2024kkb}, topological defects accompanying the outgoing states and twisted conformal boundary states~\cite{Maldacena:1995pq,vanBeest:2023dbu,Tachikawa:2026cxd,Cordova:2026dtw,Tong:2026apr}, and composite operators in Polchinski's fermion--rotor model and its extensions~\cite{Polchinski:1984uw,Loladze:2024ayk,Loladze:2025jsq,Hamada:2026jgj,Koundouros:2026rnb}.
Other directions include proposals for outgoing states beyond the single-particle $s$-wave sector retained in the two-dimensional reduction~\cite{Csaki:2021ozp}, and studies of massive fermions in monopole backgrounds, albeit still using effective core boundary conditions~\cite{Hook:2024vhf,Dawson:1983cm,Khoze:2026ueo}.
The phenomenological implications of monopole-catalyzed proton decay and the unresolved unitarity puzzle motivate a better understanding of the Rubakov--Callan process and, more broadly, monopole--fermion dynamics.

Ref.~\cite{Csaki:2024ajo} calculated the Rubakov--Callan correlator within the two-dimensional reduction of the Abelian effective field theory (EFT), showing that catalysis by an infinitely heavy monopole, represented by a 't Hooft line, is mediated by an Abelian instanton.
This instanton is a saddle point of the effective action, sourced by the fermions and carrying nonzero Abelian topological charge.
Here we extend this analysis to the spontaneously broken Georgi--Glashow $SU(2)$ theory with gauge coupling $g$ and charged gauge-boson mass $m_W$, replacing the singular defect with a smooth 't Hooft--Polyakov monopole~\cite{tHooft:1974kcl,Polyakov:1974ek}.
One qualitative difference is that at energies $g^2 m_W$ charge deposition on the monopole core becomes possible~\cite{Preskill:1984gd, Polchinski:1984uw}.
In contrast to approaches that represent charge deposition through an effective rotor~~\cite{Callan:1982ah,Callan:1982au,Goldstein:1983bk,Polchinski:1984uw,
Preskill:1984gd,Craigie:1985gk,Isler:1987xn,Affleck:1993np,
Yegulalp:1994eq,Brennan:2021ewu,Hamada:2022eiv,Brennan:2023tae,
Bogojevic:2024xtx,Loladze:2024ayk,Loladze:2025jsq,Hamada:2026jgj}, we follow Ref.~\cite{Csaki:2024ajo} and retain the dynamical gauge fields explicitly.
This allows us to calculate the electric fields generated by the fermions and identify their role in monopole catalysis.

Our main result is the identification of a novel type of instanton that mediates proton-decay catalysis in this UV completion.
We call these configurations sourced $SU(2)$ instantons because the fermions participating in the process act as sources for the dynamical gauge fields.
For the $SU(2)$ Rubakov--Callan correlator studied here, we find non-Abelian configurations with instanton number $Q=1$.
This topological index is directly related to the number of fermion insertions and is independent of their positions relative to the monopole core.
Our explicit count of the number of fermionic zero modes agrees with the expectation from the Atiyah--Singer index theorem~\cite{Atiyah:1963zz}, although this theorem cannot be directly applied here because of the monopole world line.
(Note that index theorems for static monopole backgrounds in three dimensions~\cite{Callias:1977kg,Brennan:2021ucy} do not directly apply in this four-dimensional setting.)
We find that, unlike BPST instantons, the amplitudes associated with these sourced instantons are not exponentially suppressed by $e^{-8\pi^2/g^2}$~\cite{Belavin:1975fg}.

We calculate the Rubakov--Callan correlator to leading order in the weak-coupling limit.
A closed-form analytic expression is given in Eq.~\eqref{eq:correlator solution}, and is verified numerically.
The corresponding scattering amplitude exhibits two regimes.
In the IR, where the fermions have energies $E \ll g^2m_W$, we recover the Abelian-instanton description and correlator of Ref.~\cite{Csaki:2024ajo}.
The Rubakov--Callan amplitude scales as $\mathcal M_{\rm RC}\propto E^{-1}$, independently of $g$ and $m_W$, and saturates $s$-wave unitarity.
The leading correlator is completely captured by the exterior Abelian gauge dynamics.
Suppression sets in at energies $E\sim g^2m_W$, where charge deposition on the monopole becomes possible.
For $E\gg g^2m_W$, the amplitude scales as $\mathcal M_{\rm RC}\propto\pqty{g^2m_W/E}^5E^{-1}$.
We interpret this suppression as competition with charge-deposition channels and possibly other channels that become available at higher energies.
Note that for GUT monopoles, $g^2 m_W \sim 10^{15}\, {\rm GeV}$, so all phenomenologically relevant catalysis is in the regime that saturates unitarity.

Our results indicate that a careful re-examination of monopole scattering described using effective boundary conditions or a rotor model is warranted.
The sourced instanton and its fermion zero modes are localized well outside the monopole core, at radii $r\gtrsim(g^2m_W)^{-1}$.
This indicates that the Rubakov--Callan process is governed by IR dynamics outside the core.
The scattering fermions do not reach the monopole core.
The agreement between the instanton number and the fermion zero-mode count suggests that configurations with fractional instanton number cannot supply the same zero modes required by the correlators considered here.
Since our sourced instantons can be mapped to windings of Polchinski's effective rotor~\cite{Polchinski:1984uw}, this suggests that the same set of normalizable fermion zero modes cannot contribute to semiton solutions involving partial windings of the rotor.
Additionally, we can see that the limit $m_W r \to \infty$ and $g \to 0$ is quite subtle: at fixed physical insertion separations, keeping $g^2m_W r$ finite or sending it to infinity correspond to qualitatively different regimes, even though the monopole core becomes point-like in both limits.

%%%%%%%%%%%
{\bf Setup.}---%
%%%%%%%%%%%
Consider Georgi--Glashow theory with an $SU(2)$ gauge field $A_\mu^a$, gauge coupling $g$, and a real adjoint scalar $\Phi^a$, which we refer to as the Higgs.
We use canonically normalized fields, with $D_\mu=\partial_\mu-igA_\mu^a\sigma^a/2$.
The Higgs potential $V=\lambda\pqty{\Phi^a\Phi^a-v^2}^2/4$ leads to spontaneous symmetry breaking from $SU(2)$ to $U(1)$, with charged gauge-boson mass $m_W=gv$.
In the BPS limit $\lambda \ll g^2$ considered here, the Higgs is parametrically lighter than the $W$~\cite{Bogomolny:1975de,Prasad:1975kr}.
We consider a static 't Hooft--Polyakov monopole at the origin, described by the background fields
\begin{equation}
	A_i^a=\frac{\epsilon_{aij}x_j}{r}\frac{1-K(r)}{gr}\,,
	\: A_0^a=0\,,
	\: \Phi^a=-H(r)\frac{x^a}{r}\,,
\end{equation}
where $K(r)=m_Wr/\sinh(m_Wr)$ and $H(r)=(m_Wr\coth(m_Wr)-1)/(gr)$.

To model the fermion sector relevant to the Rubakov--Callan process in an $SU(5)$ GUT, one only needs to focus on the particular $SU(2)$ subgroup of $SU(5)$ which is broken to $U(1)$ via the $SU(5)$ adjoint VEV.
(The $U(1)$ associated with the monopole embedding is distinct from the electromagnetic $U(1)_{\mathrm{em}}$.)
This is the $SU(2)$ we will be considering here, and under this $SU(2)$ the fermions of the Rubakov--Callan process are each contained in a different $SU(2)$ doublet.
Hence in our simplified $SU(2)$ model we introduce $N_f=4$ massless left-handed Weyl doublets $\Psi_f$ with an $SU(4)$ flavor symmetry and vanishing Yukawa couplings.
The four doublets are~\cite{Callan:1982ac}
\begin{eqnarray}
	\begin{pmatrix}
		e \\ -\bar{d}^3
	\end{pmatrix}\,,
	\qquad
	\begin{pmatrix}
		\bar{u}^1 \\ u^2
	\end{pmatrix}\,,
	\qquad
	\begin{pmatrix}
		-\bar{u}^2 \\ u^1
	\end{pmatrix}\,,
	\qquad
	\begin{pmatrix}
		d^3 \\ \bar{e}
	\end{pmatrix}\,,
\end{eqnarray}
where $e$, $u^i$, and $d^i$ denote left-handed electron and quark fields, and barred fields denote left-handed antiparticles.
A dagger denotes the antiparticle of the indicated field, {\it i.e.} $\bar{e}$ and $e^\dagger$ denote left- and right-handed positrons, respectively.

We are interested in the leading-order contribution to the charge-conserving flavor-changing Euclidean correlator which represents the Rubakov--Callan process
\begin{eqnarray} \label{eq:correlator}
	&&C\pqty{x_1,x_2,x_3,x_4}
	=\frac{\epsilon^{f_1f_2f_3f_4}}{4!}\\[3pt]
	&&\quad\ev{
		\Psi_{+,f_1}(x_1)
		\Psi_{+,f_2}(x_2)
		W_{13}\Psi_{-,f_3}(x_3)
		W_{24}\Psi_{-,f_4}(x_4)
	}\,.\nonumber
\end{eqnarray}
Here $\Psi_{+,f}$ and $\Psi_{-,f}$ denote the upper and lower $SU(2)$ components of $\Psi_f$ in the local charge basis, respectively, carrying charges $\pm g/2$ under the residual $U(1)$ gauge symmetry.
The insertion points are $x_k=(t_k,\vb{x}_k)$, with $r_k=\abs{\vb{x}_k}$.
The factors $W_{ij}$ are non-Abelian Wilson lines from $x_i$ to $x_j$ which ensure the gauge invariance of the correlator.

%%%%%%%%%%%
{\bf Two-dimensional reduction.}---%
%%%%%%%%%%%
To calculate this correlator, we expand the fields in eigenstates of the grand angular momentum $\vb{J}=\vb{L}+\vb{S}+\vb{T}$, where $\vb{L}$, $\vb{S}$, and $\vb{T}$ denote orbital angular momentum, spin, and isospin, respectively.
The isospin contribution accounts for the fermion--monopole pairwise helicity, which has opposite signs for the two charged components of each $SU(2)$ doublet~\cite{Csaki:2021ozp}.
Following the seminal works of Rubakov and Callan~\cite{Rubakov:1981rg,Callan:1982ac}, we retain the lowest fermion partial waves, $j=0$, which capture the leading non-perturbative contribution to monopole catalysis.
This sector contains the fermion zero modes responsible for the unitarity-saturating correlator~\cite{Csaki:2024ajo}.
The $1/r$ effective radial masses of higher fermion partial waves outside the core provide the usual heuristic motivation for the $s$-wave truncation.
However, to our knowledge, the absence of higher fermion partial waves from the leading outgoing state has not been rigorously established.
Higher-partial-wave fermion loops can also induce RG running of the coupling.
These corrections are subleading in the weak-coupling regime considered here.

Angular-momentum conservation forbids a vertex containing two $j=0$ fermions and a gauge field with $j \neq 0$, motivating the spherical truncation of the gauge sector as well.
A novel feature of our calculation is that we retain the dynamical $s$-wave gauge fields throughout, which is crucial for maintaining a gauge-invariant description.
Their importance was already highlighted in Ref.~\cite{Csaki:2024ajo}, which showed that $s$-wave photon dynamics generate the Abelian instantons responsible for monopole catalysis in the IR.

Retaining the lowest fermion partial waves and using the Witten ansatz~\cite{Witten:1976ck} for the spherically symmetric gauge fields, we integrate over the angles to obtain the reduced Euclidean action,
\begin{eqnarray}
	&&S_{\rm 2D}
	=4\pi\int \dd{t}\,\dd{r}\,\Biggl[
		\frac{r^2}{4g^2}f_{\alpha\beta}f_{\alpha\beta}
		+\frac{1}{g^2}\pqty{\partial_\alpha k}^2
		\label{eq:2D action}
		\\
	&&+\frac{k^2}{g^2}\pqty{\partial_\alpha\varphi-a_\alpha}^2
		+\frac{\pqty{1-k^2}^2}{2g^2r^2}
		+\frac{r^2}{2}\pqty{\partial_\alpha h}^2
		+k^2h^2
	\Biggr]
	\nonumber
	\\
	&&+\sum_{f=1}^{N_f}\int \dd{t}\,\dd{r}\,
	\bar\psi_f
	\bqty{
		\gamma^\alpha\pqty{\partial_\alpha-\frac{i}{2}a_\alpha\gamma_5}
		+\frac{k}{r}e^{-i\gamma_5\varphi}
	}\psi_f \,.
	\nonumber
\end{eqnarray}
Here $\alpha, \beta \in \Bqty{t, r}$ are 2D indices and the lowercase fields $a_\alpha(t,r)$, $f_{\alpha \beta}(t,r)$, $w(t,r) \equiv k(t,r) e^{i \varphi(t,r)}$, $h(t,r)$ and $\psi_f(t,r)$ correspond to the $j=0$ coefficients of the photon, photon field strength, $W$-boson, Higgs and fermions, respectively.
In the static monopole background, $k(t,r)=K(r)$ and $h(t,r)=H(r)$.

%%%%%%%%%%%
{\bf Outline of the calculation.}---%
%%%%%%%%%%%
We proceed to calculate the leading-order correlator within the reduced two-dimensional theory, Eq.~\eqref{eq:2D action}.
The relevant physical processes depend on the distance of the fermion insertions from the origin.
At distances $r\lesssim m_W^{-1}$, the fermions probe the non-Abelian fields inside the monopole core.
Charge deposition on the monopole costs an energy of order $g^2m_W$ and can compete with the Rubakov--Callan process at distances $r\sim(g^2m_W)^{-1}$, well outside the core.
At distances $r\gg(g^2m_W)^{-1}$, the monopole is effectively pointlike and we recover the Abelian limit and results of Ref.~\cite{Csaki:2024ajo}.
Below we outline the main steps of the calculation.
The full derivation is given in Appendix~\ref{app:lo:computation}.

We first integrate out the $s$-wave fermions and obtain the effective bosonic action.
In the sectors with non-trivial instanton number, $\int F^{a\mu\nu}\widetilde F^a_{\mu\nu}\neq0$, the gauge fields support fermionic zero modes, which lead to a 't~Hooft vertex and supply the key mechanism for the flavor-changing Rubakov--Callan process.
The overlap of the fermion insertions in the correlator~\eqref{eq:correlator} with these zero modes provides the only non-perturbative contribution to the correlator and generates sources for the bosonic fields.
The 2D chiral anomaly (bubble diagram) leads the photon to acquire a Schwinger mass $m_{\rm Sch}^2(r)=N_fg^2/(16\pi^2r^2)$~\cite{Schwinger:1962tp}.
Although the Schwinger mass term arises at one loop, it affects the soft photon mode~\cite{Rubakov:1981rg,Csaki:2024ajo} at leading order.
At distances $r\gg 1/(g^2m_W)$, this mode becomes pure gauge and its kinetic term $f^2$ vanishes, while its winding can still give a finite boundary contribution fixed by the chiral anomaly.
The Schwinger mass term therefore governs the IR dynamics of this soft mode and is essential for capturing the sourced instanton that underlies the Rubakov--Callan process.
The remaining fermion loops do not lead to such IR enhancement and give only perturbative corrections to the other fields.
In Appendix~\ref{app:lo:loops} we explicitly check all one-loop contributions from the $s$-wave fermions through quadratic order in the bosons and verify that only the Schwinger mass term contributes at leading order.

At leading order in $g^2$, $k(t,r)=K(r)$ and $h(t,r)=H(r)$, corresponding to the background monopole BPS profiles.
Corrections to these profiles induced by the fermion sources enter perturbatively, as shown in Appendix~\ref{app:lo:computation}.
The $W$ phase $\varphi(t,r)$ and the photon $a_\alpha(t,r)$ remain dynamical.
After subtracting the monopole self-energy contribution to the action (the terms containing only $k$ and $h$ in Eq.~\eqref{eq:2D action}), we obtain the following action in Lorenz gauge $\partial_\alpha a_\alpha=0$
\begin{eqnarray}
	&& S_{\rm 2D}^{\rm eff}
	=\frac{4\pi}{g^2}\int\dd{t}\,\dd{r}\,
	\Biggl[\frac{r^2}{4}f_{\alpha\beta}^2
	+\frac{r^2}{2}m_{\rm Sch}^2(r)a_\alpha^2
    \label{eq:photon-phi action}\\
    &&\quad+K(r)^2\pqty{\partial_\alpha\varphi - a_\alpha}^2\Biggr]+\int\dd{t}\,\dd{r}\,\pqty{j^{\alpha}a_\alpha+\rho_{\rm z.m.}^{\varphi}\varphi} \,.
	\nonumber
\end{eqnarray}
Here $j^{\alpha}$ and $\rho_{\rm z.m.}^{\varphi}$ are the sources for $a_\alpha$ and $\varphi$ generated by the fermion insertions and the Wilson lines.

Apart from capturing the dynamics of the photon, the action~\eqref{eq:photon-phi action} also describes charge deposition on the monopole.
The deposited charge produces an electric field whose energy makes the resulting dyon heavier than the neutral monopole.
For $n$ units of electric charge $g/2$, the BPS mass formula determines $M_{\rm dyon}-M_{\rm mon}\simeq n^2g^2m_W/(32\pi)$~\cite{Preskill:1984gd}, identifying $g^2m_W$ as the charge-deposition scale.
This energy cost is reflected in the coefficients of the first and third terms in Eq.~\eqref{eq:photon-phi action}.

Varying the action in Eq.~\eqref{eq:photon-phi action} gives coupled, inhomogeneous equations for $a_\alpha$ and $\varphi$.
To solve them we also need to specify the boundary conditions.
At large Euclidean distances, the photon field $a_\alpha$ falls as $1/\sqrt{t^2+r^2}$.
This allows configurations that approach a pure gauge with non-trivial winding at infinity, contributing to the correlator when the appropriate fermionic sources, such as those in the correlator~\eqref{eq:correlator}, are included~\cite{Csaki:2024ajo}.

The second boundary condition is at the origin.
To determine it for the photon, we consider the electric energy of charge deposited on the monopole.
Concentrating a nonzero charge at $r=0$ would give $f_{tr}\propto r^{-2}$ and make the integral of the electric term $r^2f_{tr}^2/(2g^2)$ in Eq.~\eqref{eq:2D action} diverge.
We therefore require $r^2f_{tr}\to0$ at the origin.
The magnetic term $(1-k^2)^2/(2g^2r^2)$ likewise makes the action diverge unless $k(t,0)=1$.
Since the Yukawa couplings vanish, the fermions do not source the Higgs directly.
The correction induced through $k$ is subleading, so the Higgs follows its BPS profile at leading order, with $h(t,0)=H(0)=0$.
We fix the gauge-dependent boundary value of $\varphi$ to the time-independent constant $\varphi(t,0)=0$ modulo $2\pi$, so $\partial_t\varphi(t,0)=0$.
Since $k(t,0)=1$, a nonzero $a_t(t,0)$ would make the electric-field component $k(\partial_t\varphi-a_t)/r$ diverge as $1/r$.
We therefore require $a_t(t,0)=0$.

The four-dimensional fermion fields and their first derivatives must be finite and continuous at the origin.
These conditions make the two-dimensional components vanish at least linearly, $\psi_{\pm,f}\propto r$.
The resulting boundary conditions are $\psi_{\pm,f}(t,0)=0$ and $\partial_r\psi_{+,f}(t,0)=\partial_r\psi_{-,f}(t,0)$.
(In other gauges, the second condition is modified to $\partial_r\psi_{+,f}(t,0)=e^{i\varphi(t,0)}\partial_r\psi_{-,f}(t,0)$.)

%%%%%%%%%%%
\begin{figure*}[ht!]
	\centering
	\includegraphics[width=\textwidth]{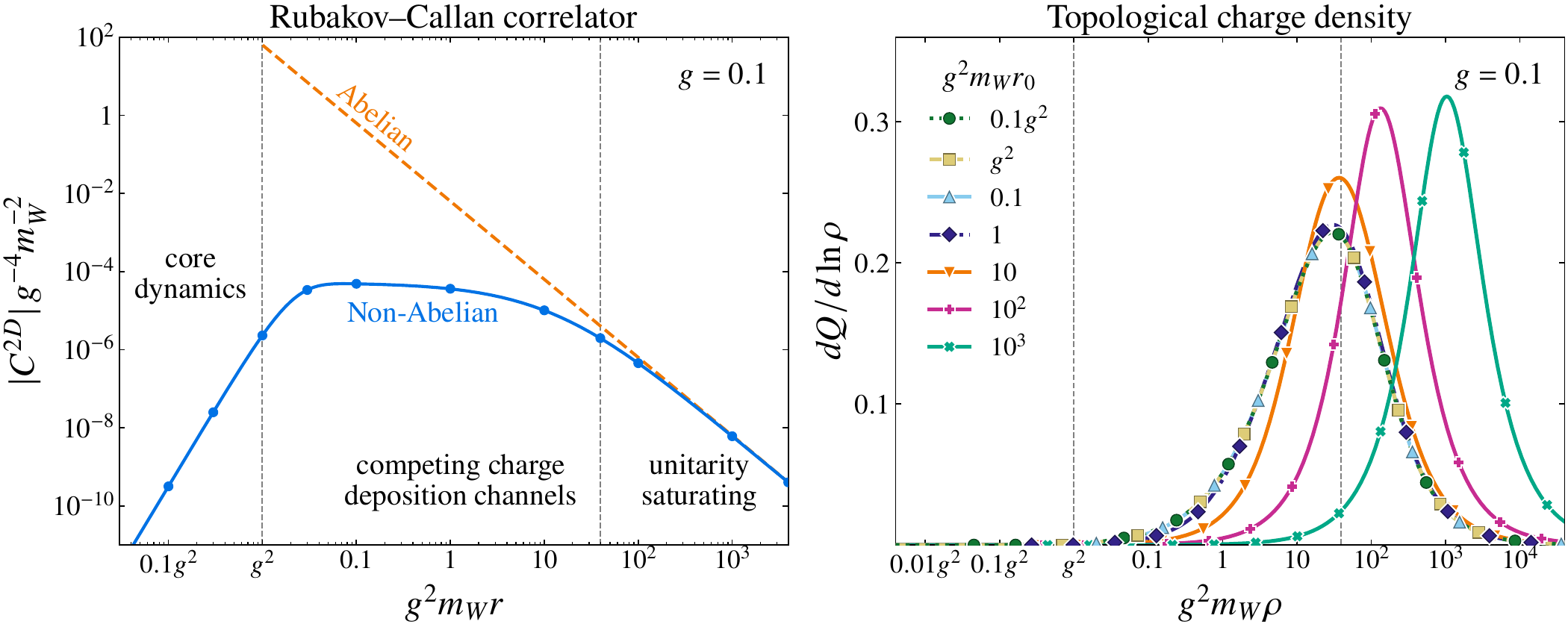}
	\caption{
		\textbf{Rubakov--Callan correlator and topological charge.}
		\textit{Left:} Radial correlator at $r_i=r$ and $t_i=0$ (blue), compared with the Abelian point-monopole result (orange dashed), where the markers denote numerical results.
		It scales as $r^4$ near the origin, approaches a plateau for $1\ll m_Wr\ll g^{-2}$, and recovers the Abelian result in the IR.
		The gray dashed lines mark the core radius $m_Wr=1$ and the charge-deposition scale $g^2m_Wr=4\pi^2$.
		\textit{Right:} Topological charge density as a function of the four-dimensional radius $\rho=\sqrt{r^2+t^2}$, for insertions at $t_i=0$ and $r_i=r_0$, with the different curves corresponding to $g^2m_Wr_0=0.1g^2,g^2,0.1,1,10,10^2,10^3$.
		Markers denote numerical results.
		Both analytic and numerical distributions integrate to unity.
		The gray dashed lines mark $m_W\rho=1$ and $g^2m_W\rho=4\pi^2$.
	}
	\label{fig:correlator-and-topology}
\end{figure*}
%%%%%%%%%%%

%%%%%%%%%%%
{\bf Results.}---%
%%%%%%%%%%%
The leading-order two-dimensional correlator is given by
\begin{eqnarray}
    &&C^{2D}=\prod_{k=1}^4\tanh\frac{m_Wr_k}{2}
    \prod_{\substack{i=1,2\\j=3,4}}G\bqty{g^2m_W(z_j+\bar z_i)}\,,
    \nonumber\\[5pt]
    &&G(\xi)\equiv\sqrt{\frac{g^2m_W}{2\pi\xi}}\,
    \exp\bqty{-\frac12e^{\xi/(4\pi^2)}E_1\pqty{\frac{\xi}{4\pi^2}}}\,,
    \label{eq:correlator solution}
\end{eqnarray}
where $z_i=r_i+it_i$, $t_i$ is Euclidean time, and $E_1$ is the exponential integral.
We can recover the leading four-dimensional correlator~\eqref{eq:correlator} by restoring the angular profiles of the zero modes,
\begin{eqnarray}
	C=C^{2D}\prod_{i=1}^2\frac{\mathcal Y_+(\theta_i,\phi_i)}{r_i}
	\prod_{j=3}^4\frac{\mathcal Y_-(\theta_j,\phi_j)}{r_j}\,,
	\label{eq:correlator-reconstruction}
\end{eqnarray}
where the expressions for the spinors $\mathcal Y_\pm$ are provided in Appendix~\ref{app:lo:conventions}.

The left panel of Fig.~\ref{fig:correlator-and-topology} shows the radial correlator for equal-time insertions at a common radius, $t_i=t$ and $r_i=r$.
We have verified this result by numerically solving the saddle-point equations, finding agreement up to corrections of order $g^2$.
For fermion insertions at radii $g^2m_Wr_i\gg1$, the characteristic energies probed by the correlator, $E_i\sim r_i^{-1}$, lie below the charge-deposition scale.
In this regime, the leading-order correlator reduces to the Abelian-instanton result of Ref.~\cite{Csaki:2024ajo} and saturates $s$-wave unitarity.
LSZ reduction of the correlator (see Appendix~\ref{app:radial-lsz}) yields a scattering amplitude scaling as $\mathcal M_{\rm RC}\propto E^{-1}$, which is independent of both $g$ and $m_W$.

At smaller radii, $m_W^{-1}\ll r\ll(g^2m_W)^{-1}$, the correlator approaches its peak value $e^{2\gamma_E}g^4m_W^2/(64\pi^6)$.
We interpret this suppression relative to the IR as evidence for competition between charge deposition and the Rubakov--Callan channel in which the monopole remains neutral.
For example, with the incoming state $u^1+u^2+M$, a final $\bar u^1+\bar u^2$ pair would require $D_{-2}$, corresponding to a dyon with charge $-2g$.
At $r\ll m_W^{-1}$, the factor $\tanh^4(m_Wr/2)$ makes the correlator vanish as $O(r^4)$.
We attribute this further suppression to competition with additional channels involving the monopole core.
In both cases, the associated scattering amplitude, obtained by LSZ reduction, scales as $\mathcal M_{\rm RC}\propto\pqty{g^2m_W/E}^5E^{-1}$ ($E\gg g^2m_W$).

For a GUT monopole, typical values $m_W\sim10^{16}$~GeV and $g^2\sim0.3$ give a charge-deposition scale of $g^2m_W/(4\pi^2)\sim10^{14}$~GeV.
The hadronic energies relevant to proton decay lie far below this scale, in the regime where the Rubakov--Callan channel saturates the unitarity bound.
At fixed coupling, the short-distance suppression we find outside the monopole core would become relevant at hadronic energies only if the symmetry-breaking scale were much lower.
Our results therefore  do not affect existing bounds on GUT monopoles from proton decay searches~\cite{Super-Kamiokande:2012tld,MACRO:2002iaq,IceCube:2014xnp}.

%%%%%%%%%%%
{\bf Sourced $SU(2)$ Instantons.}---%
%%%%%%%%%%%
The essential ingredient in our calculation has been the novel type of gauge field configuration which we call the sourced $SU(2)$ instanton.
As in the Abelian case~\cite{Csaki:2024ajo}, we obtain this saddle for the gauge fields by solving the equations of motion in the presence of the fermion sources.
This sourced instanton carries a non-trivial winding number (Pontryagin index).
However, unlike ordinary instantons, its contribution to the path integral is not suppressed by the usual $e^{-8\pi^2/g^2}$ instanton factor, and it also does not satisfy the (anti)self-dual condition for the conventional BPST instanton.
The reason behind these unusual properties of the sourced instanton is the presence of the static monopole background.
The topological-charge distribution is shown in the right panel of Fig.~\ref{fig:correlator-and-topology}.
We have verified numerically that the distribution agrees with the analytical result and that both integrate to unit topological charge, irrespective of the positions of the fermion insertions.
We can in fact easily show that this topological quantum number must be an integer.
The reduction of the 4D winding  number $Q\propto\int F\tilde{F}$ to 2D is still a total derivative, and hence will be a combination of boundary terms
\begin{align}
    2\pi Q&=\int_{\mathbb{H}}\dd^2x\,\epsilon^{\alpha\beta}\partial_\alpha V_\beta
    \nonumber\\
    &=-\int_\mathbb{R}\dd{t}\,V_t\bigg|^{r=\infty}_{r=0} +\int_0^\infty\dd{r}\,V_r\bigg|^{t=\infty}_{t=-\infty}\,,
    \label{eqn: topo-density}
\end{align}
where $V_\beta=\big(1-K^2\big)(a_\beta-\partial_\beta\varphi)$\,, $\epsilon^{tr}=\epsilon_{tr}=1$\,, and $\mathbb H$ denotes the right half plane.
Staticity in the asymptotic past and future eliminates the second term.
The contribution at the origin drops out as $K(0) = 1$.
The remaining term can be rewritten using the equations of motion as
\begin{equation}\label{eqn:Q-count}
  Q=\frac{1}{2\pi}\int_{\mathbb{H}}\mathrm{d}^2x\,
    \epsilon^{\alpha\beta}\partial_\alpha a_\beta
   =\frac{2n_{A}}{N_f}\,.
\end{equation}
Here $n_A\equiv-\int_{\mathbb{H}}\mathrm{d}^2x\,\epsilon^{\alpha\beta} \partial_\alpha j_{\beta}$ counts the total axial charge inserted, to which each fermion insertion contributes $1/2$.
For the Rubakov--Callan process, with $N_f=4$ and four insertions, our counting gives exactly $Q=1$.

An important aspect of these instantons is that they are sourced by the fermionic insertions in the correlator.
The leading-order contribution to the correlator comes from the 't Hooft vertex, which has one leg for each of the fermion zero modes.
We now explain how the sourced instanton indeed does admit the right number of fermion zero modes.
A more detailed treatment is given in Appendix~\ref{app: quant-charge}.
For our non-compact half plane, a fermionic zero mode $\psi$ is a normalizable (or at least log-normalizable) solution to the Weyl equation which is regular both at the origin and at infinity.
For each flavor, the regularity at infinity restricts the class of solutions to polynomially decaying functions $\psi\sim|r+i t|^{n-Q}$ labelled by $n=0,1,2\dots$, where $Q$ is the topological charge given by \eqref{eqn:Q-count}.
(Here we assume that $Q$ is a positive integer.
For negative $Q$, the zero modes instead occur in $\bar\psi$.)
For each $n$, there are two zero-mode solutions.
The boundary condition at the origin $r=0$ eliminates one, leaving a unique solution for each $n$.

The requirement of log-normalizability truncates the tower at $n\leq\lfloor Q-1\rfloor$.
Therefore, we have exactly $Q$ zero modes for each flavor.
For $Q=1$ and $N_f=4$, this gives exactly the number of fermion insertions in the correlator~\eqref{eq:correlator}.
This shows that our calculation is self-consistent.
 
What happens to correlators with a different number of fermionic insertions?
Suppose for example that we keep  $N_f=4$ but with only two fermions inserted in the correlator.
In this case, Eq.~\eqref{eqn:Q-count} gives $Q=1/2$.
Hence we could have expected to see two zero modes to source this {\it half instanton} configuration.
However, log-normalizability requires $n\leq\lfloor-1/2\rfloor=-1$, meaning that there are no normalizable zero modes.
Hence there is no appropriate 't Hooft vertex matching the number of insertions in the correlator, and thus the correlator vanishes in the weak-coupling limit.

Note that the gauge part of the full action in Eq.~\eqref{eq:2D action} satisfies the Bogomolny bound $S_{\rm YM}\geq8\pi^2|Q|/g^2$~\cite{Bogomolny:1975de}.
Over a Euclidean time interval $T$, the static BPS monopole contributes $S_{\rm mon}=4\pi m_WT/g^2$, half of which comes from the Yang--Mills field~\cite{Prasad:1975kr}.
For $m_WT\gg4\pi$, as assumed here, this magnetic contribution alone already exceeds $8\pi^2/g^2$.
The effective bosonic action in Eq.~\eqref{eq:photon-phi action} subtracts this common monopole self-energy, leaving an excess action that is not subject to the same constraint.
Indeed, we can directly calculate the change in the effective action and see that $\Delta S\ll g^{-2}$ both inside and outside the core in the weak-coupling limit.
Taking all insertions at the same point $t_i=t$ and $r_i=r$, we find
\begin{eqnarray}
	\Delta S=2e^{\frac{g^2m_Wr}{2\pi^2}}
	E_1\pqty{\frac{g^2m_Wr}{2\pi^2}}
	{}-4\ln\tanh\frac{m_Wr}{2}\,.
\end{eqnarray}
Inside the core, $m_Wr\ll1$, we find $\Delta S\sim-\ln(g^2m_W^3r^3)$.
The action grows as the insertions approach the origin, suppressing the correlator.
Outside the core, at $m_W^{-1}\ll r\ll(g^2m_W)^{-1}$, we find $\Delta S\sim-\ln(g^2m_Wr)$, which diverges only logarithmically as $g\to0$.
In the IR, at $r\gg(g^2m_W)^{-1}$, $\Delta S\simeq4\pi^2/(g^2m_Wr)\to0$.

The topological-charge distribution in Fig.~\ref{fig:correlator-and-topology} reveals another important feature.
The topological charge and the corresponding fermion zero-mode profiles remain localized well outside the monopole core, at radii $r\gtrsim(g^2m_W)^{-1}$, even when the fermion insertions lie \emph{inside} the core.
This indicates that the sourced instanton is an IR effect and that the participating scattering fermions do not reach the core.

%%%%%%%%%%%
{\bf Relation to the rotor model.}---%
%%%%%%%%%%%
An alternative approach to the one taken in this manuscript describes charge deposition on the monopole as excitation of a quantum-mechanical rotor~\cite{Callan:1982ah,Callan:1982au,Goldstein:1983bk,Polchinski:1984uw,
Preskill:1984gd,Craigie:1985gk,Isler:1987xn,Affleck:1993np,
Yegulalp:1994eq,Brennan:2021ewu,Hamada:2022eiv,Brennan:2023tae,
Bogojevic:2024xtx,Loladze:2024ayk,Loladze:2025jsq,Hamada:2026jgj}.
We compare our findings with those of the rotor model.
The rotor $\theta (t)$ is related to the photon $a_\alpha$ and the $W$-boson phase $\varphi$, and is assumed to capture the entire dynamics of these fields.
The precise connection is made by first introducing the field  $\theta(t,r)=\varphi(t,r)+\int_r^\infty\dd r'\,a_r(t,r')$, and then performing a gauge transformation with the gauge transformation parameter $\lambda=\theta-\varphi$.
This sets the gauge fields $a_r'=0$ and $\varphi'=\theta$, while $a_t'=a_t+\partial_t\lambda$.
In the rotor model one retains $\theta(t)=\theta(t,0)$ as a dynamical variable while dropping the remaining photon field $a_t'$ from the dynamical degrees of freedom.

The same gauge transformation gives $\psi'_{\pm,f}=e^{\pm i\lambda/2}\psi_{\pm,f}$, with $\psi'_{\pm,f}(t,0)=0$ and $\partial_r\psi'_{+,f}(t,0)=e^{i\theta(t)}\partial_r\psi'_{-,f}(t,0)$.
Refs.~\cite{Kazama:1983xp,Kazama:1983rt,Lam:1984sm,Yamagishi:1984zu,Brennan:2021ewu,Hamada:2022eiv,Brennan:2023tae,Loladze:2024ayk,Loladze:2025jsq} replace this boundary condition with the effective relation $\psi^{\rm ext}_{+,f}(t,0)=e^{i\theta(t)}\psi^{\rm ext}_{-,f}(t,0)$ for fermions that cannot resolve the core.
The effective origin represents the core of radius $r\sim m_W^{-1}$, while the exterior fields describe distances $r\gg m_W^{-1}$ at energies $E\ll m_W$.

After integrating out the fermions, the quadratic Euclidean rotor action is
\begin{eqnarray}\label{rotor}
	S_{\rm rotor}^{(2)}
	&=&\frac I2\int\dd t\,\dot\theta^2
	+\frac{N_f}{16\pi^2}\int\dd t\,\dd t'\,
	\frac{\bqty{\theta(t)-\theta(t')}^2}{(t-t')^2} \,, \nonumber \\[5pt] 
\end{eqnarray}
where $I=4\pi/(g^2m_W)$ is the moment of inertia, which accounts for the monopole charge-deposition energy.
The non-local term is the anomaly contribution associated with the Schwinger mass term in Eq.~\eqref{eq:photon-phi action}.
Calculating the correlator in the rotor model with this effective boundary condition reproduces the functional form of Eq.~\eqref{eq:correlator solution} at $r_i\gg m_W^{-1}$.
In this sense, the effective boundary condition replaces the 't Hooft vertex of the sourced instanton.

We can further identify the unit instanton number with the $2\pi$ winding of Polchinski's rotor associated with the Rubakov--Callan process~\cite{Loladze:2025jsq}, $Q=\Delta\theta/(2\pi)=1$.
Fractional windings $0<\Delta\theta<2\pi$ correspond to $0<Q<1$, for which our calculation finds no normalizable fermion zero modes.
Such fractional changes of the rotor angle occur in the fermion--rotor description proposed as a resolution of the unitarity puzzle~\cite{Loladze:2024ayk,Loladze:2025jsq}.
Our zero-mode count therefore implies that the same fermion zero modes responsible for the Rubakov--Callan correlator cannot participate in those processes.
Since the effective boundary condition appears to implement the sourced instanton's contribution to the Rubakov--Callan process, one should be cautious about using it to resolve the unitarity puzzle.

%%%%%%%%%%%
{\bf Outlook.}---%
%%%%%%%%%%%
Our study of the Rubakov--Callan effect reveals a new type of topological configuration, which we call a \emph{sourced instanton}.
The agreement between the number of fermion zero modes and the instanton number suggests an underlying index theorem, although we have not found an existing theorem that directly applies to this setup.

Our results indicate that Rubakov--Callan catalysis is an IR effect in which the scattering fermions do not reach the monopole core.
The sourced instanton and the fermion zero modes are localized well outside it.

Our calculation of monopole catalysis in the $SU(2)$ Georgi--Glashow theory provides a starting point for studying perturbative and radiative corrections.
Extending this calculation would, for example, allow the effects of renormalization-group running to be investigated systematically.

More broadly, the sourced instantons found here and in Ref.~\cite{Csaki:2024ajo} provide a QFT implementation of quantum selection rules in monopole-catalyzed proton decay.
It would be interesting to investigate whether sourced instantons can realize selection rules in other systems.

\hspace{5pt}

%%%%%%%%%%%
\noindent\emph{Note added.---}%
%%%%%%%%%%%
During the final stages of preparing this manuscript, Ref.~\cite{Geng:2026xob} appeared, which also studies the UV completion of fermion--monopole scattering.

\hspace{5pt}

\begin{acknowledgments}
%%%%%%%%%%%
\noindent{\bf Acknowledgements.}---%
%%%%%%%%%%%
	We thank John Terning and Marius Kongsore for useful discussions.
	We also thank Ofri Telem and John Terning for valuable comments on our manuscript.
	CC and LK are supported in part by the NSF grant PHY-2609553 and in part by the US-Israeli BSF grant 2024091.
	ZL is supported by NSF grant PHY-2409350.
	RO acknowledges partial support from the ERC Starting Grant “Light-Dark” (Grant No. 101040019), the Milner Fellowship, and the Azrieli Fellowship.
	RO also thanks Cornell University for its gracious hospitality.
	This project has received funding from the European Research Council (ERC) under the European Union’s Horizon Europe research and innovation programme (grant agreement No. 101040019).
	Views and opinions expressed are however those of the author(s) only and do not necessarily reflect those of the European Union.
	The European Union cannot be held responsible for them.
\end{acknowledgments}

\appendix

\setcounter{secnumdepth}{3}

\renewcommand{\thesection}{\Alph{section}}
\renewcommand{\thesubsection}{\thesection.\arabic{subsection}}
\renewcommand{\thesubsubsection}{\thesubsection.\arabic{subsubsection}}

\makeatletter
\renewcommand{\p@subsection}{}
\renewcommand{\p@subsubsection}{}
\makeatother

\onecolumngrid

%%%%%%%%%%%
\section{Calculation of the two-dimensional correlator}
\label{app:lo:computation}
%%%%%%%%%%%

In this appendix, we present a detailed semiclassical calculation of the leading Rubakov--Callan correlator at weak coupling, using a 't~Hooft--Polyakov monopole in the BPS limit.
Our calculation applies to fermion insertions both inside and outside the monopole core.
Appendix~\ref{app:lo:conventions} gives the two-dimensional reduction, and Appendix~\ref{app:lo:integration} performs the fermion integral for the $N_f$ left-handed Weyl doublets.
Appendix~\ref{app:lo:loops} calculates the one-loop corrections to the bosonic action.
Appendix~\ref{app:lo:zeromodes} gives the fermion zero modes in a bosonic background and their contributions to the sources.
Appendix~\ref{app:lo:fieldequations} derives the bosonic equations of motion and their boundary conditions and shows that $k$ and $h$ reduce to the static BPS profiles at leading order.
Appendix~\ref{app:lo:greens} gives the leading saddle-point solution for the photon $\sigma$ and the $W$-field phase $\varphi$, together with the corresponding fermion zero-mode profiles.
Appendix~\ref{app:lo:correlator} evaluates the correlator and fixes its normalization.

%%%%%%%%%%%
\subsection{Two-dimensional reduction}
\label{app:lo:conventions}
%%%%%%%%%%%

We start from four-dimensional $SU(2)$ Georgi--Glashow theory with $N_f$ massless left-handed Weyl doublets and no Yukawa couplings.
We absorb the gauge coupling into $A_\mu^a$, so $A_\mu^a/g$ is the canonically normalized gauge field.
The Euclidean action is
\begin{eqnarray}
	S_{\rm 4D}&=&\int\dd[4]{x}\,\Biggl[
	\frac{1}{4g^2}F^a_{\mu\nu}F^a_{\mu\nu}
	+\frac{1}{2}(D_\mu\Phi)^a(D_\mu\Phi)^a
	+\frac{\lambda}{4}\pqty{\Phi^a\Phi^a-v^2}^2
	+\sum_{f=1}^{N_f}\bar\Psi_f\pqty{D_t+i\sigma^iD_i}\Psi_f
	\Biggr]\,,
	\label{app:lo:4daction}
\end{eqnarray}
where $\Psi_f$ is the Weyl fermion of flavor $f$, $\Phi^a$ is the adjoint Higgs field, $v$ is its vev, and $g$ and $\lambda$ are the gauge and Higgs self-couplings.
The Pauli matrices $\sigma^i$ act on spin indices.
The covariant derivatives are $D_\mu\Psi_f=(\partial_\mu-iA_\mu^aT^a)\Psi_f$, with $T^a=\tau^a/2$, and $(D_\mu\Phi)^a=\partial_\mu\Phi^a+\epsilon^{abc}A_\mu^b\Phi^c$.
We take $N_f$ even to avoid the $SU(2)$ Witten anomaly~\cite{Witten:1982fp,Loladze:2025jsq}.
Ultimately we take $N_f=4$ to calculate the correlator~\eqref{eq:correlator}.

The Higgs vev spontaneously breaks $SU(2)$ to $U(1)$ and gives the charged gauge bosons mass $m_W=gv$.
In the BPS limit $\lambda/g^2\to0$, the static monopole \cite{Bogomolny:1975de,Prasad:1975kr} is
\begin{eqnarray}
	A_i^a=\frac{\epsilon_{aij}x_j}{r^2}\bqty{1-K(r)}\,,\qquad
	A_t^a=0\,,\qquad \Phi^a=-H(r)\frac{x^a}{r}\,,
	\label{app:lo:monopolebackground}
\end{eqnarray}
where $r=\abs{\vb x}$ denotes the radial spatial coordinate.
The profiles are
\begin{eqnarray}
	K(r)=\frac{m_Wr}{\sinh(m_Wr)}\,,\qquad
	H(r)=\frac{m_Wr\coth(m_Wr)-1}{gr}\,.
	\label{app:lo:bps}
\end{eqnarray}
In the IR, the Rubakov--Callan process involves the $j=0$ fermion partial wave~\cite{Rubakov:1981rg,Callan:1982ac,Csaki:2024ajo}.
The monopole is invariant under combined spatial and isospin rotations, so the conserved angular momentum is $\vb J=\vb L+\vb S+\vb T$, where $\vb L$, $\vb S$, and $\vb T$ denote orbital angular momentum, spin, and isospin~\cite{Jackiw:1976xx}.
For an $SU(2)$ doublet, this isospin contribution accounts for the radial angular momentum associated with pairwise helicity~\cite{Csaki:2021ozp}.

We expand in eigenstates of $\vb J^2$ and $J_z$.
In the $j=0$ sector, spin and isospin combine to $\Sigma=0,1$, with orbital angular momentum $l=\Sigma$.
The corresponding angular wavefunctions are
\begin{eqnarray}
	\mathcal Y_0=\frac{1}{\sqrt{8\pi}}
	\begin{pmatrix}
		0&1\\
		-1&0
	\end{pmatrix}\,,\qquad
	\mathcal Y_1=(\vb*{\sigma}\cdot\vu r)\mathcal Y_0
	=\frac{1}{\sqrt{8\pi}}
	\begin{pmatrix}
		-\sin\theta\,e^{-i\phi}&\cos\theta\\
		\cos\theta&\sin\theta\,e^{i\phi}
	\end{pmatrix}\,.
	\label{app:lo:angularharmonics}
\end{eqnarray}
The row index is spin and the column index is isospin, with normalization $\int\dd\Omega\,\operatorname{tr}(\mathcal Y_A^\dagger\mathcal Y_B)=\delta_{AB}$.
Each four-dimensional Weyl field therefore has the expansion
\begin{eqnarray}
	\Psi_f(t,\vb x)=\frac{1}r\bqty{u_f(t,r)\mathcal Y_0
	+v_f(t,r)\mathcal Y_1}+\text{higher partial waves}\,.
	\label{app:lo:weylexpansion}
\end{eqnarray}
The unbroken $U(1)$ charge generator is $T_r=-\vu r\cdot\vb T$.
Its eigenstates are $\mathcal Y_\pm=(\mathcal Y_0\pm\mathcal Y_1)/\sqrt2$, with eigenvalues $\pm1/2$.
Their radial coefficients $\psi_{\pm,f}=(u_f\pm v_f)/\sqrt2$ thus carry physical charges $\pm g/2$.

At tree level, two $j=0$ fermions couple only to a $j=0$ gauge field.
The Witten ansatz~\cite{Witten:1976ck} parameterizes these $j=0$ gauge fields and contains the static monopole as a background.
Including the spherical Higgs field, we write
\begin{eqnarray}
	A_t^a=-a_t \hat r_a\,,\qquad \Phi^a=-h \hat r_a\,,\qquad
	A_i^a=\frac{(1-k\cos\varphi)\epsilon_{aij}\hat r_j
	-k\sin\varphi(\delta_{ai}-\hat r_a\hat r_i)}{r}
	-a_r \hat r_a\hat r_i\,.
	\label{app:lo:sphericalansatz}
\end{eqnarray}
Here $a_\alpha(t,r)$, with $\alpha\in\{t,r\}$, is the radial photon, $w(t,r)=k(t,r)e^{i\varphi(t,r)}$ is the charged $W$ field, and $h(t,r)$ is the Higgs profile.
The monopole background has $a_\alpha=\varphi=0$\,, $k=K$\,, and $h=H$\,.
Under a gauge transformation $U_\xi=\exp[-i\xi(t,r)\hat r_aT^a]$, $a_\alpha\mapsto a_\alpha+\partial_\alpha\xi$ and $w\mapsto e^{i\xi}w$, so $D_\alpha w=(\partial_\alpha-ia_\alpha)w$ and $f_{\alpha\beta}=\partial_\alpha a_\beta-\partial_\beta a_\alpha$.

After integrating over the angles and taking the BPS limit $\lambda/g^2\to0$, we obtain the bosonic action
\begin{eqnarray}
	S_B&=&4\pi\int\dd t\,\dd r\,\Biggl[
	\frac{r^2}{4g^2}f_{\alpha\beta}f_{\alpha\beta}
	+\frac{(\partial_\alpha k)^2+k^2(\partial_\alpha\varphi-a_\alpha)^2}{g^2}
	+\frac{(1-k^2)^2}{2g^2r^2}
	+\frac{r^2}{2}(\partial_\alpha h)^2+k^2h^2\Biggr]\,.
	\label{app:lo:angularcontractions}
\end{eqnarray}
Note that a two-dimensional gauge field can be written in terms of two scalar functions,
\begin{eqnarray}
	a_\alpha=\partial_\alpha\upsilon
	-\epsilon_{\alpha\beta}\partial^\beta\sigma\,,\qquad \epsilon_{tr}=1\,.
	\label{app:lo:gaugedecomposition}
\end{eqnarray}
We choose $\upsilon=0$, which imposes Lorenz gauge $\partial_\alpha a_\alpha=0$, allowing us to represent the photon by a scalar $\sigma$,
\begin{eqnarray}
	a_t=-\partial_r\sigma\,,\qquad a_r=\partial_t\sigma\,,\qquad
	f_{tr}=\partial^2\sigma\,,\qquad
	\partial^2=\partial_t^2+\partial_r^2\,.
	\label{app:lo:hodge}
\end{eqnarray}
We use $a_\alpha$ and $\sigma$ interchangeably when referring to the photon.
In $S_B$, the $W$ phase appears through the gauge-invariant combination
\begin{eqnarray}
	\mathcal A_\alpha\equiv\partial_\alpha\varphi-a_\alpha
	=\partial_\alpha\varphi+\epsilon_{\alpha\beta}\partial^\beta\sigma\,,
	\qquad
	\partial_\alpha\mathcal A_\beta-\partial_\beta\mathcal A_\alpha
	=-f_{\alpha\beta}\,.
	\label{app:lo:invariant}
\end{eqnarray}

For fermions in the lowest partial wave, we define $\psi_f=(\psi_{+,f},\psi_{-,f})^T$.
With $\gamma^t=\sigma_2$\,, $\gamma^r=-\sigma_1$\,, and $\gamma_5=-i\gamma^t\gamma^r=\sigma_3$, the reduced action is
\begin{eqnarray}
	S_{\rm 2D}=S_B+\sum_f\int\dd t\,\dd r\,
	\bar\psi_f\slashed{D}\psi_f+\text{higher partial waves}\,,\qquad
	\slashed{D}\equiv
	\gamma^\alpha\pqty{\partial_\alpha-\frac{i}{2}a_\alpha\gamma_5}
	+\frac{k}{r}e^{-i\gamma_5\varphi}\,.
	\label{app:lo:reducedweyl}
\end{eqnarray}
In the region $m_W^{-1}\ll r\ll(g^2m_W)^{-1}$, the Kaluza--Klein (KK) decomposition gives the higher fermion partial waves a mass term $\sqrt{j(j+1)}/r$, with $j\geq1$~\cite{Csaki:2024ajo}.
The conventional argument for suppressing the higher partial waves is that these masses suppress their amplitudes near the monopole core~\cite{Callan:1982ac}.
Here we give a different argument based on their IR behavior.
For the regular solutions of the point-monopole exterior theory, at fixed $t$ and $r\gg(g^2m_W)^{-1}$ (and far from all fermion insertion points), we find
\begin{eqnarray}
	\frac{\abs{\psi_j}}{\abs{\psi_{j=0}}}
	\propto r^{-\sqrt{j(j+1)}}\longrightarrow0
	\qquad(r\to\infty\,,\ j\geq1)\,,
	\label{app:lo:higherwavefalloff}
\end{eqnarray}
where $\psi_j$ is the radial coefficient of the $j$th fermion partial wave.
We note that higher-partial-wave fermion loops can lead to RG running~\cite{Csaki:2024ajo}, which is a perturbative correction in the regime studied here.
We estimate the relative correction to the coupling to be of order $N_fg^2\ln(m_W/\mu)/(16\pi^2)$, where $\mu\ll m_W$ is of order the fermion energy scale.
At $\mu\sim g^2m_W$ this is $O(N_fg^2\abs{\ln g^2}/(16\pi^2))$, which vanishes in the $g\to0$ limit.
We therefore neglect this running at leading order.
To the best of our knowledge, there is no rigorous argument for strictly excluding higher partial waves from fermion--monopole scattering (Ref.~\cite{Csaki:2021ozp} provides a symmetry-based argument for why they may be necessary in the out-state of the unitarity puzzle).
We leave a more detailed quantitative investigation of these higher partial waves to future work.

%%%%%%%%%%%
\subsection{Integrating out the fermions}
\label{app:lo:integration}
%%%%%%%%%%%

To calculate the correlator, we seek a saddle point for the gauge and Higgs fields in the presence of fermion insertions.
The fermion insertions source the gauge fields and can therefore lead them to deviate from their background BPS profiles.
However, the saddle-point equations couple the fermions and bosons, so the fermion profiles themselves depend on the bosonic profiles.
We therefore first integrate out the fermions assuming a general bosonic background $a_\alpha,k,\varphi,h$, following Ref.~\cite{Csaki:2024ajo}, and then solve for the sourced fields.

We denote the $n_0$ normalizable zero modes per flavor by $\psi_a^{(0)}$, with $\slashed{D}\psi_a^{(0)}=0$.
Each Weyl field can be decomposed into the zero-mode contributions and the nonzero-mode contributions (denoted with a prime),
\begin{eqnarray}
	\psi_f(x)=\sum_{a=1}^{n_0}c_{fa}\psi_a^{(0)}(x)+\psi_f'(x)\,.
	\label{app:lo:modesplit}
\end{eqnarray}
The barred field has the analogous expansion in $\bar n_0$ zero modes per flavor satisfying $\bar\psi_b^{(0)}\slashed{D}=0$.
(Note that $\bar n_0$ is not necessarily equal to $n_0$.)
We introduce fermion sources $\bar\eta_f,\eta_f$ through $\int\dd[2]{x}\,(\bar\eta_f\psi_f+\bar\psi_f\eta_f)$.
Integration over the coefficients $c_{fa}$ and their barred counterparts, followed by Gaussian integration over the nonzero modes, gives
\begin{eqnarray}
	Z_F[\bar\eta,\eta;a_\alpha,k,\varphi,h]
	&=&\pqty{\det'\slashed{D}}^{N_f}\mathcal P_0[\bar\eta,\eta]
	\exp\bqty{\sum_f\int\dd[2]{x}\,\dd[2]{y}\,
	\bar\eta_f(x)G'(x,y)\eta_f(y)}\,,\nonumber\\
	\mathcal P_0&=&\prod_{f=1}^{N_f}
	\bqty{\prod_{a=1}^{n_0}\int\dd[2]{x}\,\bar\eta_f(x)\psi_a^{(0)}(x)}
	\bqty{\prod_{b=1}^{\bar n_0}\int\dd[2]{x}\,\bar\psi_b^{(0)}(x)\eta_f(x)}\,.
	\label{app:lo:fermionintegral}
\end{eqnarray}
Integrating over the zero modes gives the factor $\mathcal P_0$, which is the 't~Hooft vertex.
The exponential contains the propagator $G'(x,y)$ for the nonzero modes.
The determinant accounts for quantum fermion-loop corrections.
All three factors depend on the bosonic background.

The partition function for the bosonic fields is a sum over topological sectors with different zero-mode structures,
\begin{eqnarray}
	Z[\bar\eta,\eta]=\sum_{\text{topological sectors}}
	\int\mathcal D a_\alpha\,\mathcal D k\,\mathcal D\varphi\,\mathcal D h\,
	e^{-S_B}Z_F[\bar\eta,\eta;a_\alpha,k,\varphi,h]\,.
	\label{app:lo:partition}
\end{eqnarray}
In Appendix~\ref{app: quant-charge} we relate these sectors to the instanton number of the gauge configurations.

The four-dimensional Wilson line along a path $\gamma_{ij}$ from $x_i$ to $x_j$ is
\begin{eqnarray}
	W_{ij}[A]=\mathcal P\exp\bqty{i\int_{\gamma_{ij}}
	A_\mu^aT^a\dd{x^\mu}}\,,
	\label{app:lo:Wilson4D}
\end{eqnarray}
where $\mathcal P$ denotes path ordering.
In the monopole background, the charged $W$ contribution $K/r=m_W/\sinh(m_Wr)$ decays exponentially outside the core.
For the correlator~\eqref{eq:correlator}, we choose paths in the $(t,r)$ plane at fixed spatial angles, oriented from the positive-charge insertion $x_i$ to the negative-charge insertion $x_j$.
The leading correlator depends only on the endpoints of these paths in the $(t,r)$ plane.
Since $\dd{x^i}=\hat r_i\dd r$, the angular terms in Eq.~\eqref{app:lo:sphericalansatz} give
\begin{eqnarray}
	\epsilon_{aij}\hat r_i\hat r_j=0\,,\qquad
	(\delta_{ai}-\hat r_a\hat r_i)\hat r_i=0\,,\qquad
	A_\mu^aT^a\dd{x^\mu}=T_r(a_t\dd t+a_r\dd r)\,.
	\label{app:lo:Wilsoncontraction}
\end{eqnarray}
The generator $T_r$ is constant along each path, so path ordering is no longer needed.
Projecting onto the charge components in the correlator gives
\begin{eqnarray}
	W_{ij}[a]=\exp\bqty{\frac{i}{2}\int_{\gamma_{ij}}
	(a_t\dd t+a_r\dd r)}\,,\qquad J_{\rm WL}=W_{13}W_{24}\,.
	\label{app:lo:Wilsonlines}
\end{eqnarray}
Thus $\delta\ln J_{\rm WL}/\delta\varphi= \delta\ln J_{\rm WL}/\delta k=0$.
The Wilson-line contribution to the photon current is
\begin{eqnarray}
	j_{\rm WL}^{\alpha}(x)
	=-\frac{\delta\ln J_{\rm WL}}{\delta a_\alpha(x)}
	=-\frac{i}{2}\sum_{\gamma=\gamma_{13},\gamma_{24}}
	\int_\gamma\dd y^\alpha\,\delta^{(2)}(x-y)\,.
	\label{app:lo:Wilsoncurrent}
\end{eqnarray}
Thus $J_{\rm WL}=\exp[-\int\dd[2]{x}\,j_{\rm WL}^\alpha a_\alpha]$, and the generating functional with the Wilson lines is
\begin{eqnarray}
	Z_W[\bar\eta,\eta]=\sum_{\text{topological sectors}}
	\int\mathcal D a_\alpha\,\mathcal D k\,\mathcal D\varphi\,\mathcal D h\,
	e^{-S_B-\int\dd[2]{x}\,j_{\rm WL}^\alpha a_\alpha}
	Z_F[\bar\eta,\eta;a_\alpha,k,\varphi,h]\,.
	\label{app:lo:Wilsonpartition}
\end{eqnarray}

For $N_f=4$, we use $x_i=(t_i,r_i)$ and $z_i=r_i+it_i$ for the radial insertion points.
Differentiating with respect to the four fermion sources gives
\begin{eqnarray}
	C^{2D}(x_1,x_2,x_3,x_4)
	=\frac{\epsilon^{f_1f_2f_3f_4}}{4!\,Z[0,0]}
	\left.
	\frac{\delta^4 Z_W[\bar\eta,\eta]}
	{\delta\bar\eta_{+,f_1}(x_1)\,
	\delta\bar\eta_{+,f_2}(x_2)\,
	\delta\bar\eta_{-,f_3}(x_3)\,
	\delta\bar\eta_{-,f_4}(x_4)}
	\right|_{\eta=\bar\eta=0}\,.
	\label{app:lo:correlatorderivative}
\end{eqnarray}
Restoring the angular wavefunctions and the four factors $1/r_i$ gives the $j=0$ contribution to $C$, as displayed in Eq.~\eqref{eq:correlator-reconstruction}.

In the sector with four zero modes (one per flavor) and no barred zero modes, the four insertions saturate the 't~Hooft vertex.
The resulting process can saturate unitarity, and does so in the IR~\cite{Csaki:2024ajo}.
For $n_0=1$ and $\bar n_0=0$, the product of normalized zero-mode wavefunctions is
\begin{eqnarray}
	J_0=\psi_+^{(0)}(x_1)\psi_+^{(0)}(x_2)
	\psi_-^{(0)}(x_3)\psi_-^{(0)}(x_4)\,.
	\label{app:lo:zeromodefactor}
\end{eqnarray}
(For logarithmically normalizable modes, the divergent normalization factors cancel against inverse factors from $\det'\slashed D$~\cite{Csaki:2024ajo} and can thus be omitted from $J_0$ below.)
Including the Wilson lines, the photon current and the sources for the $W$ phase and amplitude are
\begin{eqnarray}
	j^{\alpha}(x)
	=-\frac{\delta\ln J_0}{\delta a_\alpha(x)}+j_{\rm WL}^\alpha(x)\,,\qquad
	\rho_{\rm z.m.}^{\varphi}(x)=-\frac{\delta\ln J_0}{\delta\varphi(x)}\,,\qquad
	\rho_{\rm z.m.}^{k}(x)=-\frac{\delta\ln J_0}{\delta k(x)}\,.
	\label{app:lo:insertionforces}
\end{eqnarray}
For the photon $\sigma$, $a_\alpha=-\epsilon_{\alpha\beta}\partial^\beta\sigma$ gives the scalar source
\begin{eqnarray}
	\rho^{\sigma}(x)\equiv\epsilon_{\alpha\beta}\partial^\beta
	j^{\alpha}(x)\,.
	\label{app:lo:scalarsource}
\end{eqnarray}
%

%%%%%%%%%%%
\subsection{One-loop quantum corrections}
\label{app:lo:loops}
%%%%%%%%%%%

As we will show in the following appendices, loop effects can be important in monopole--fermion scattering.
This happens when infrared enhancement compensates for their coupling suppression.
In this appendix, we calculate the one-loop contributions to the effective bosonic action arising from $\det'\slashed D$ in Eq.~\eqref{app:lo:fermionintegral}.

The photon vacuum-polarization diagram gives the contribution from the 2D chiral anomaly
\begin{eqnarray}
	\vcenter{\hbox{%
		\colorlet{bubblecolor}{.}%
		\begin{tikzpicture}[
			x=1mm,y=1mm,font=\small,text=bubblecolor,
			fermion/.style={draw=bubblecolor,line width=.65pt},
			photon/.style={draw=bubblecolor,line width=.65pt,decorate,
				decoration={snake,amplitude=.85pt,segment length=4.4pt}},
			vertex/.style={circle,fill=bubblecolor,inner sep=1.15pt}
		]
			\draw[fermion,postaction={decorate},decoration={markings,
				mark=at position .25 with {\arrow[rotate=-10]{Stealth[length=1.6mm,width=1.1mm]}},
				mark=at position .75 with {\arrow[rotate=-10]{Stealth[length=1.6mm,width=1.1mm]}}}]
				(0,0) circle (4);
			\draw[photon] (-10,0)--(-4,0);
			\draw[photon] (4,0)--(10,0);
			\node[vertex] at (-4,0) {};
			\node[vertex] at (4,0) {};
			\node[left,inner sep=1.8pt] at (-10,0) {$a_\alpha$};
			\node[right,inner sep=1.8pt] at (10,0) {$a_\beta$};
		\end{tikzpicture}%
	}}\ \Longrightarrow\quad
	W_{\rm anom}&=&\frac{N_f}{8\pi}\int\dd t\,\dd r\,
	a_\alpha\bqty{\delta_{\alpha\beta}-\frac{\partial_\alpha\partial_\beta}{\partial^2}}a_\beta
	=\frac{N_f}{8\pi}\int\dd t\,\dd r\,(\partial_\alpha\sigma)^2\,,
	\label{app:lo:scales}
\end{eqnarray}
where in the second equality we use the Lorenz gauge $\partial_\alpha a^\alpha=0$.
This transverse quadratic term gives the gauge-invariant Schwinger mass~\cite{Schwinger:1962tp} $m_{\rm Sch}^2(r)=N_fg^2/(16\pi^2r^2)$.

The Schwinger mass term contributes at leading order because the photon has a soft mode.
Using Rubakov's decomposition~\cite{Rubakov:1982fp}, we write $\sigma=\eta+\mathcal F$ outside the monopole core.
Away from the fermion sources, the quadratic photon action takes the form
\begin{eqnarray}
	S_{\gamma}^{(2)}
	&=& \int\dd{t}\,\dd{r}\,\bqty{
		\frac{2\pi}{g^2}r^2\pqty{\partial^2\mathcal F}^2
		+\frac{N_f}{8\pi}\pqty{\partial_\alpha\pqty{\eta+\mathcal F}}^2
	}.
\end{eqnarray}
Away from the sources, the gauge potential of the soft mode $\eta$ becomes locally pure gauge as $\partial^2\eta=0$.
The electric field is then determined entirely by $\mathcal F$ through $f_{tr}=\partial^2\sigma=\partial^2\mathcal F=m_{\rm Sch}^2(r)\mathcal F$.
Consider the Abelian instanton vortex of Ref.~\cite{Csaki:2024ajo}, sourced by fermion insertions at radius $r_0$ and time $t=0$.
At $\rho=\sqrt{t^2+r^2}\gg r_0$, its gauge potential is $a_\alpha\simeq-2\epsilon_{\alpha\beta}\partial^\beta\ln(\rho/r_0)$, with $\partial_\rho\eta\simeq 2/\rho$ and $\mathcal F=O(rr_0/\rho^2)$.
The soft mode thus contributes $(N_f/8\pi)(\partial_\alpha\eta)^2\simeq N_f/(2\pi\rho^2)$ to the Schwinger action density.
The mixed term falls as $O(r_0/\rho^3)$, while both the $(\partial_\alpha\mathcal F)^2$ term and the Maxwell density $(2\pi/g^2)r^2f_{tr}^2$ fall as $O(r_0^2/\rho^4)$.
The Schwinger term therefore governs the IR dynamics of the photon and is essential for capturing the sourced instanton underlying the Rubakov--Callan process.
In fact, at insertion radii $r_0\gg 1/(g^2m_W)$, the energy required to deposit charge on the monopole exceeds the characteristic fermion energy $E\sim 1/r_0$, and the soft photon gives the dominant contribution to the four-point function~\eqref{eq:correlator solution}.

To calculate the remaining loop corrections, we first determine the Feynman rules.
We expand about $k=K$\,, $\varphi=0$\,, and $a_\alpha=0$, writing $\delta k=k-K$.
The position-space fermion propagator on the BPS background is
\begin{eqnarray}
	\radialpropdiagram=G^{\prime}_{\rm BPS}(t-t';r,r')
	=\bqty{-\gamma^t\partial_t-\gamma^r\partial_r+\frac{K(r)}{r}}
	\bqty{\frac{1+\gamma^r}{2}g_+(t-t';r,r')+\frac{1-\gamma^r}{2}g_-(t-t';r,r')}\,,
	\label{app:lo:loopscattering}
\end{eqnarray}
where
{\small
\begin{eqnarray}
	g_+(t-t';r,r')&=&\frac1\pi
	\pqty{\partial_r-\frac{m_W}{2}\tanh\frac{m_Wr}{2}}
	\pqty{\partial_{r'}-\frac{m_W}{2}\tanh\frac{m_Wr'}{2}}
	\int_0^\infty\dd\omega_E\,
	\frac{e^{-\omega_E\abs{t-t'}}\cos(\omega_Er)\cos(\omega_Er')-e^{-2\omega_E/m_W}}
	{\omega_E(\omega_E^2+m_W^2/4)}\,,
	\nonumber\\
	g_-(t-t';r,r')&=&\frac1\pi
	\pqty{\partial_r-\frac{m_W}{2}\coth\frac{m_Wr}{2}}
	\pqty{\partial_{r'}-\frac{m_W}{2}\coth\frac{m_Wr'}{2}}
	\int_0^\infty\dd\omega_E\,
	\frac{e^{-\omega_E\abs{t-t'}}\sin(\omega_Er)\sin(\omega_Er')}
	{\omega_E(\omega_E^2+m_W^2/4)}\,.
	\label{app:lo:radialkernels}
\end{eqnarray}
}
The subtraction in $g_+$ fixes an infrared constant that drops out of $G^{\prime}_{\rm BPS}$.
The vertices are
\begin{eqnarray}
	\begin{array}{ccc}
		\radialonevertex{dashed}{\delta k}\;=\;\dfrac{1}{r}
		&\qquad\radialonevertex{photon}{a_t}\;=\;-\dfrac{i\gamma^t\gamma_5}{2}
		&\qquad\radialonevertex{photon}{a_r}\;=\;-\dfrac{i\gamma^r\gamma_5}{2}
		\\[22pt]
		\radialonevertex{dashed}{\varphi}\;=\;-\dfrac{iK}{r}\gamma_5
		&\qquad\radialtwovertex{\varphi}{\varphi}\;=\;-\dfrac{K}{r}
		&\qquad\radialtwovertex{\delta k}{\varphi}\;=\;-\dfrac{i\gamma_5}{r}\,.
	\end{array}
	\label{app:lo:loopvertices}
\end{eqnarray}
There is no direct Higgs vertex because the Yukawa couplings vanish.

Apart from the Schwinger mass, the nonvanishing loops give, up to quadratic order,
\begin{eqnarray}
	\Delta S_{\rm ferm}
	&=&\frac{4\pi}{g^2}\Biggl[-\frac{N_fg^2}{4\pi}\,\radiallooptadpole{\delta k}
	+\frac{N_fg^2}{8\pi}\,\radialloopbubble{\delta k}{\delta k}
	\nonumber\\[5pt]
	&&\hspace{34pt}-\frac{N_fg^2}{8\pi}\,\radialloopcontact{\varphi}{\varphi}
	+\frac{N_fg^2}{8\pi}\,\radialloopbubble{\varphi}{\varphi}
	+\cdots\Biggr]\,,
	\label{app:lo:loopcomplete}
\end{eqnarray}
where the ellipsis denotes cubic and higher terms.
Under charge conjugation, $\delta k$ remains unchanged while $\varphi$ and $a_\alpha$ change sign.
The mixed $\delta k a_\alpha$ and $\delta k\varphi$ diagrams, as well as the photon and phase tadpoles, therefore vanish.
The mixed $a_\alpha\varphi$ contribution vanishes in Lorenz gauge.

We calculate loop corrections in the $\overline{\mathrm{MS}}$ scheme with a renormalization scale $\mu$.
The tadpole diagram evaluates to
\begin{eqnarray}
	-\left.\frac{\delta\Delta S_{\rm ferm}}{\delta k(t,r)}\right|_{\delta k=\varphi=a_\alpha=0}
	=N_f\,\radiallooptadpole{\delta k}
	=\frac{N_f}{\pi r}\bqty{\frac{K}{r}\pqty{\frac{\operatorname{Ei}(m_Wr)+\operatorname{Ei}(-m_Wr)}{2}+\ln\frac{\mu}{m_W}}-\frac1{2r}}\,,
	\label{app:lo:moduluslooptadpole}
\end{eqnarray}
where $\operatorname{Ei}$ is the exponential integral.
The quadratic one-loop contributions are given by
\begin{eqnarray}
	\Pi_k(t-t';r,r')=\radialloopbubble{\delta k}{\delta k}
	=\frac{1}{rr'}\operatorname{tr}\bqty{G^{\prime}_{\rm BPS}(t-t';r,r')G^{\prime}_{\rm BPS}(t'-t;r',r)}\,,
	\label{app:lo:generalkkernel}
\end{eqnarray}
\begin{eqnarray}
	\Pi_\varphi(t-t';r,r')&=&-\radialloopcontact{\varphi}{\varphi}+\radialloopbubble{\varphi}{\varphi}
	\nonumber\\[5pt]
	&=&\frac{K(r)}{\pi r^2}\bqty{K(r)\pqty{\frac{\operatorname{Ei}(m_Wr)+\operatorname{Ei}(-m_Wr)}{2}+\ln\frac{\mu}{m_W}}-\frac12}\delta(t-t')\delta(r-r')
	\nonumber\\[5pt]
	&&-\frac{K(r)K(r')}{rr'}\operatorname{tr}\bqty{G^{\prime}_{\rm BPS}(t-t';r,r')\gamma_5G^{\prime}_{\rm BPS}(t'-t;r',r)\gamma_5}\,.
	\label{app:lo:generalphasekernel}
\end{eqnarray}
At large distances $\sqrt{(t-t')^2+r^2}\gg m_W^{-1}$, we find that
\begin{eqnarray}
	\Pi_k(t-t';r,r')\simeq
	-\frac{2}{\pi^2\bqty{(t-t')^2+(r-r')^2}\bqty{(t-t')^2+(r+r')^2}}\,,\qquad r,r'\gg m_W^{-1}\,,
	\label{app:lo:generalloopIR}
\end{eqnarray}
which in momentum space approaches a finite value in the IR.
The other quadratic terms in $\Delta S_{\rm ferm}$ decay exponentially outside the core, at $r\gg m_W^{-1}$.
The terms in $\Delta S_{\rm ferm}$ therefore contribute only at subleading order in $g^2$ and can be dropped in our leading-order analysis.
Their UV logarithms are associated with the RG running of the parameters of the theory.

With $S_B$ from Eq.~\eqref{app:lo:angularcontractions}, the leading sourced action used to calculate the correlator is then
\begin{eqnarray}
	S_{\rm 2D}^{\rm eff}=S_B+\int\dd t\,\dd r\,\bqty{\frac{N_f}{8\pi}(\partial_\alpha\sigma)^2+\rho^{\sigma}\sigma+\rho_{\rm z.m.}^{\varphi}\varphi+\rho_{\rm z.m.}^{k}\delta k}\,.
	\label{app:lo:effectiveaction}
\end{eqnarray}
%

%%%%%%%%%%%
\subsection{Fermion zero modes}
\label{app:lo:zeromodes}
%%%%%%%%%%%

The zero mode fermions solve the equation
\begin{eqnarray}
	\slashed{D}\psi_f^{(0)}
	=\left[\gamma^\alpha\left(\partial_\alpha-\frac{i}{2}a_\alpha\gamma_5\right)
	+\frac{k}{r}e^{-i\gamma_5\varphi}\right]
	\begin{pmatrix}\psi^{(0)}_{+,f}\\ \psi^{(0)}_{-,f}\end{pmatrix}=0\,,
	\qquad f=1,\ldots,N_f\,.
	\label{app:lo:weylbackgroundprofile}
	\label{app:lo:dirac}
\end{eqnarray}
The origin $r=0$ is not a boundary of the four-dimensional theory, so the fermion fields and their first derivatives must be finite and continuous there.
In the gauge where $\varphi(t,0)=0$, this gives the boundary conditions
\begin{eqnarray}
	\psi^{(0)}_{+,f}(t,0)=\psi^{(0)}_{-,f}(t,0)=0\,,\qquad
	\partial_r\psi^{(0)}_{+,f}(t,0)=\partial_r\psi^{(0)}_{-,f}(t,0)\,.
	\label{app:lo:weyloriginbc}
\end{eqnarray}

We can write the solution to Eq.~\eqref{app:lo:weylbackgroundprofile} with the boundary conditions~\eqref{app:lo:weyloriginbc} in terms of the gauge fields as
\begin{eqnarray}
	\begin{pmatrix}\psi^{(0)}_{+,f}\\ \psi^{(0)}_{-,f}\end{pmatrix}
	=e^{-\sigma(t,r)/2}
	\left[1+G'_{\rm BPS}\,\frac{k e^{-i\gamma_5\varphi}-K}{r}\right]^{-1}
	\left[\exp\!\left(-\int_r^\infty\dd s\,\frac{K(s)}{s}\right)
	\begin{pmatrix}1\\1\end{pmatrix}\right]\,,
	\label{app:lo:staticweylprofile}
\end{eqnarray}
where $K(r)$ is the static BPS profile and $G'_{\rm BPS}$ is the propagator given in Eq.~\eqref{app:lo:loopscattering}.

The corresponding zero-mode contribution to the photon current is given by
\begin{eqnarray}
	\rho^\sigma_{\rm z.m.}(x)=\frac12\sum_i\delta^{(2)}(x-x_i)\,,
	\label{app:lo:source}
\end{eqnarray}
where $i$ runs over the insertion points.
Similarly, the currents for $k$ and $\varphi$ are
\begin{eqnarray}
	\begin{pmatrix}\rho^k_{\rm z.m.}(x)\\ \rho^\varphi_{\rm z.m.}(x)\end{pmatrix}
	=\frac1r\sum_i\sum_{s=\pm1}
	\frac{e^{-is\varphi(x)}\chi_{s,f_i}(x)}{\chi_{s_i,f_i}(x_i)}
	G'_{s_i s}(x_i,x)
	\begin{pmatrix}1\\ -is k(x)\end{pmatrix}\,,
	\label{app:lo:weylbackgroundsources}
\end{eqnarray}
where $s_i$ and $f_i$ label the charge component and flavor at $x_i$, and $\chi_{s,f}=e^{\sigma/2}\psi^{(0)}_{s,f}$.
The propagator in Eq.~\eqref{app:lo:weylbackgroundsources} is evaluated at $a_\alpha=0$ for the given $k$ and $\varphi$,
\begin{eqnarray}
	\left.G'\right|_{a_\alpha=0}
	=\frac{1}{(G'_{\rm BPS})^{-1}+(k e^{-i\gamma_5\varphi}-K)/r}\,.
	\label{app:lo:weylbackgroundpropagator}
\end{eqnarray}
%

%%%%%%%%%%%
\subsection{Equations of motion}
\label{app:lo:matching}
\label{app:lo:fieldequations}
\label{app:lo:profileequations}
%%%%%%%%%%%

To find the correlator~\eqref{app:lo:correlatorderivative}, we must first find the saddle point of the effective sourced action~\eqref{app:lo:effectiveaction}.
Varying with respect to $\sigma, \varphi, k$ and $h$, we find the semiclassical equations of motion for the bosonic fields
\begin{subequations}
    \begin{align}
        -\partial^2k+k(\mathcal A_\alpha\mathcal A^\alpha+g^2h^2)
        -\frac{k(1-k^2)}{r^2}
        &=-\frac{g^2}{8\pi}\rho_{\rm z.m.}^{k}\,,
        \label{app:lo:moduluseom}\\[5pt]
        \partial_\alpha(r^2\partial^\alpha h)
        &=2k^2h\,,
        \label{app:lo:higgseom}\\[5pt]
        \partial^2(r^2\partial^2\sigma)
        -2\bqty{\partial_r(k^2\mathcal A_t)-\partial_t(k^2\mathcal A_r)}
        -\frac{N_f g^2}{16\pi^2}\partial^2\sigma
        &=-\frac{g^2}{4\pi}\rho^{\sigma}\,,
        \label{app:lo:sigmaeom}\\[5pt]
        \partial_\alpha(k^2\mathcal A^\alpha)
        &=\frac{g^2}{8\pi}\rho_{\rm z.m.}^{\varphi}\,.
        \label{app:lo:phaseeom}
    \end{align}
\end{subequations}
To solve these equations of motion, we must determine the boundary conditions.
The reduced two-dimensional theory has two apparent boundaries, one at $\sqrt{r^2 + t^2}\to \infty$ and the other at the origin $r=0$.

The boundary at the origin $r=0$ is not a true boundary.
These fields $\sigma, \varphi, k$ and $h$ are partial waves of finite and smooth four-dimensional fields, and since $r=0$ is not a boundary in the four-dimensional theory, we must require they be finite and smooth at the origin.
Recall from Eq.~\eqref{app:lo:sphericalansatz} that
\begin{equation*}
    A_i^a=-\frac{k\sin\varphi}{r}\delta_{ai}
    +\frac{1-k\cos\varphi}{r}\epsilon_{aij}\hat r_j
    +\left(\frac{k\sin\varphi}{r}-a_r\right)\hat r_a\hat r_i\,.
\end{equation*}
For $A_i^a$ to be single-valued at the origin $r=0$, the coefficients of $\epsilon_{aij}\hat r_j$ and $\hat r_a\hat r_i$ must tend to zero.
Similarly, the fields $A_t^a$ and $\Phi^a$, both proportional to $\hat r^a$, must vanish there.
All told, these constraints give the conditions
\begin{equation}
    k(t,0)=1\,,\qquad h(t,0)=\varphi(t,0)=0\,,\qquad
    \partial_r\sigma(t,0)=\mathcal A_r(t,0)=0\,.
    \label{app:lo:innerbc}
\end{equation}
A point charge at the origin $r=0$ would give $f_{tr}\propto r^{-2}$, so the electric energy in Eq.~\eqref{eq:2D action} would diverge as $\int_0\dd r/r^2$.
Finite energy therefore requires $r^2 f_{tr}\to0$ as $r\to0$ (we assume that all fermion insertions lie away from the origin, $r_i > 0$).
At the second boundary, $m_W r\to\infty$, the Higgs approaches its symmetry-breaking vacuum and the massive charged $W$ field decays:
\begin{equation}
    h(t,r\to\infty)\mathrel{=}v=m_W/g\,,\qquad
    k(t,r\to\infty)\mathrel{=}0\,,\qquad f_{tr}(t,r\to\infty)=0\,.
    \label{app:lo:outerbc}
\end{equation}

The insertion source $\rho^k_{\rm z.m.}$ in Eq.~\eqref{app:lo:moduluseom} carries a factor $g^2/(8\pi)$ and therefore contributes only at subleading order, $O(N_f g^2)$ (as does the one-loop tadpole in Eq.~\eqref{app:lo:moduluslooptadpole}, which we already dropped).
The term $k\mathcal A_\alpha\mathcal A^\alpha$ is a relative $O(N_f^2 g^4)$ correction, as suggested by Eq.~\eqref{app:lo:phaseeom}, and is thus also dropped.
At leading order, Eqs.~\eqref{app:lo:moduluseom} and~\eqref{app:lo:higgseom} reduce to the source-free monopole equations,
\begin{equation}
    -\partial^2 k+k\left(g^2h^2+\frac{k^2-1}{r^2}\right)=0\,,\qquad
    \partial_\alpha(r^2\partial^\alpha h)=2k^2h\,,
    \label{app:lo:radialbackground}
\end{equation}
and thus $k$ and $h$ reduce to the static BPS monopole profiles
\begin{equation}
    k(t,r)=K(r)=\frac{m_Wr}{\sinh(m_Wr)}\,,\qquad
    h(t,r)=H(r)=\frac{1}{g}\left(m_W\coth(m_Wr)-\frac1r\right)\,.
    \label{app:lo:profilesLO}
\end{equation}

We estimate that the fermion insertions shift the BPS profiles in Eq.~\eqref{app:lo:profilesLO} by $\delta k=k-K=O(N_f g^2)$ and $g\,\delta h=g(h-H)=O(N_f g^2m_W)$.
However, we note that some fermion sources could drive the core profiles far from their BPS values.
One can verify that these estimates hold for the zero-mode source $\rho^k_{\rm z.m.}$ obtained from the solution in Appendix~\ref{app:lo:greens}, as long as all fermion insertions are taken away from the origin $r=0$.

%%%%%%%%%%%
\subsection{Saddle point solution}
\label{app:lo:greens}
\label{app:lo:split}
%%%%%%%%%%%

With the boundary conditions~\eqref{app:lo:innerbc} and~\eqref{app:lo:outerbc} and the sources~\eqref{app:lo:scalarsource} and~\eqref{app:lo:weylbackgroundsources}, we can now solve Eqs.~\eqref{app:lo:sigmaeom} and~\eqref{app:lo:phaseeom} for the photon $\sigma$ and the $W$-field phase $\varphi$.
The calculation is somewhat involved, so we quote the leading-order result for $g \ll 1$.
For fermion insertions at $(t_i,r_i)$, the leading photon field is
\begin{eqnarray}
	\sigma(t,r)
	\simeq-\frac{4}{N_f}\sum_{i=1}^{N_f}
	\ln\!\pqty{\sqrt{\frac{\pi N_f}{2m_W}}\,
	G\!\pqty{\frac{N_f g^2 m_W}{4}\bigl[r+r_i+\mathrm{i}\,s_i(t-t_i)\bigr]}}
	+\text{constant}\,,
	\label{app:lo:sigmasoft}
\end{eqnarray}
where $s_i=\pm1$ is the sign of the charge of the $i$th insertion and $G$ is defined in Eq.~\eqref{eq:correlator solution}.
The $W$-field phase is
\begin{eqnarray}
	\varphi(t,r)
	=-\frac{4i}{N_f}\sum_{i=1}^{N_f}s_i
	\ln\!\left(
		\frac{G\!\pqty{\frac{N_f g^2 m_W}{4}[r+r_i+\mathrm{i}\,s_i(t-t_i)]}}
		     {G\!\pqty{\frac{N_f g^2 m_W}{4}[r_i+\mathrm{i}\,s_i(t-t_i)]}}
	\right)\,,
	\label{app:lo:thetaposition}
\end{eqnarray}
where we have continued $\partial_r\varphi=\partial_t\sigma$ outside the core.
At leading order $k=K$, and we can set $e^{-i\gamma_5\varphi}=1$ in Eq.~\eqref{app:lo:staticweylprofile} because $\varphi$ is small where $K/r$ becomes large, and vice versa.
Thus the zero-mode profiles are
\begin{eqnarray}
	\psi^{(0)}_{+,f}(t,r)\simeq\psi^{(0)}_{-,f}(t,r)
	\simeq e^{-\sigma(t,r)/2}\tanh\frac{m_Wr}{2}\,,
	\qquad f=1,\ldots,N_f\,.
	\label{app:lo:zeromodeprofile}
\end{eqnarray}
%

%%%%%%%%%%%
\subsection{Correlator}
\label{app:lo:correlator}
%%%%%%%%%%%

With $k$ and $h$ set to their BPS profiles in Eq.~\eqref{app:lo:profilesLO}, the effective action is quadratic in the photon $\sigma$ and the $W$-field phase $\varphi$.
We can calculate the correlator by evaluating the action~\eqref{app:lo:effectiveaction} (including the source terms) on the saddle-point solutions~\eqref{app:lo:sigmasoft} and~\eqref{app:lo:thetaposition}.
In the IR limit, where $r_i\gg 1/(g^2m_W)$, the photon action reduces to that of the Abelian theory.
Following the calculation described in Ref.~\cite{Csaki:2024ajo}, we find the two-dimensional correlator
\begin{eqnarray}
	C^{2D}_{\rm IR}
	=\frac{1}{4\pi^2}\prod_{i=1}^2\prod_{j=3}^4(z_j+\bar z_i)^{-1/2}\,,
	\label{app:lo:IRnormalization}
\end{eqnarray}
where $z_i=r_i+it_i$ and $\bar z_i=r_i-it_i$.
Equation~\eqref{app:lo:IRnormalization} agrees with Refs.~\cite{Rubakov:1982fp,Callan:1982ah,Callan:1982au,Craigie:1984pc,Polchinski:1984uw,Affleck:1993np,Maldacena:1995pq,Loladze:2024ayk,Loladze:2025jsq}.
Apart from serving as a consistency check, the IR calculation allows us to determine the overall normalization constant of the correlator (see Ref.~\cite{Csaki:2024ajo} for details).
For general fermion insertion points with $r_i>0$, the leading-order correlator is
\begin{eqnarray}
	C^{2D}(x_1,x_2,x_3,x_4)
	=\prod_{k=1}^4\tanh\frac{m_Wr_k}{2}
	\prod_{i=1}^2\prod_{j=3}^4G\pqty{g^2m_W(z_j+\bar z_i)}
	\bqty{1+O\pqty{N_f g^2\ln(N_f g^2)}}\,,
	\label{app:lo:fourresult}
\end{eqnarray}
where $G$ is defined in Eq.~\eqref{eq:correlator solution}.
At short distances, where $g^2m_W\lvert z_j+\bar z_i\rvert\ll1$, we find
\begin{eqnarray}
	C^{2D}\longrightarrow
	\frac{e^{2\gamma_E}(g^2m_W)^2}{64\pi^6}
	\prod_{k=1}^4\tanh\frac{m_Wr_k}{2}\,.
	\label{app:lo:corelimit}
\end{eqnarray}
For $r_1=r_2=r_3=r_4$ and $t_1=t_2=t_3=t_4$, we numerically solved the full set of radial equations of motion~\eqref{app:lo:moduluseom}--\eqref{app:lo:phaseeom} with the sources in Eqs.~\eqref{app:lo:scalarsource} and~\eqref{app:lo:weylbackgroundsources} and found that the radial dependence of the correlator agrees with Eq.~\eqref{app:lo:fourresult} to about $1\%$ at $g=0.1$, consistent with a relative $O(g^2)$ correction.

%%%%%%%%%%%
\section{Quantization of the topological charge}
\label{app: quant-charge}
%%%%%%%%%%%
%
As shown in the main text, the topological charge of the sourced instanton is fixed by the total axial charge $n_A$ of the fermion insertions to be $Q=2n_A/N_f$, where each $\psi$ ($\bar{\psi}$) insertion contributes $1/2$ ($-1/2$).
In Appendix~\ref{app: quant-charge-computation}, we derive this relation from the equations of motion.
In Appendix~\ref{app: quant-charge-zm}, we complete the argument by showing that the number of log-normalizable fermionic zero modes is determined by the instanton number.

%%%%%%%%%%%
\subsection{Direct calculation of the topological charge}
\label{app: quant-charge-computation}
%%%%%%%%%%%

We directly calculate the $SU(2)$ instanton number $Q$ in the saddle point found in Appendix~\ref{app:lo:greens}.
This topological charge is given by
\begin{eqnarray}\label{eq:app:Q4D}
	Q&=& \frac{1}{32\pi^2}\int \dd[4]{x}\,F^{a\mu\nu}\widetilde F^a_{\mu\nu} = \frac{1}{2\pi}\int_{\mathbb H}\dd t\,\dd r\,
	\epsilon^{\alpha\beta}\partial_\alpha
	\bqty{(1-K^2)(a_\beta-\partial_\beta\varphi)}\,, 
\end{eqnarray}
where $F^{a \mu\nu}$ is the four-dimensional field strength.
In the second equality, we substituted the leading-order solution $k = K(r)$ and integrated over the angles.
The integral runs over the half-plane $\mathbb H=\Bqty{(t,r)\mid t\in\mathbb R,\ r\geq0}$.

The remaining integrand in Eq.~\eqref{eq:app:Q4D} corresponds to the contraction of the monopole magnetic field with the non-Abelian electric field induced by the fermions.
It receives a contribution from the photon $f\equiv\epsilon^{\alpha\beta}\partial_\alpha a_\beta$ and an additional contribution involving the $W$ field, given by $-\epsilon^{\alpha\beta}\partial_\alpha\bqty{K^2(a_\beta-\partial_\beta\varphi)}$.
From the presentation in Eq.~\eqref{eq:app:Q4D} it is apparent that the integrand is the curl of a 2D vector $V_\beta=(1-K^2)(a_\beta-\partial_\beta\varphi)$.
Integrating this derivative gives the boundary terms in Eq.~\eqref{eqn: topo-density},
\begin{eqnarray}
	2\pi Q &=&\int_{\mathbb H}\dd t\,\dd r\,
	\epsilon^{\alpha\beta}\partial_\alpha V_\beta 
    =-\left.\int_{\mathbb R}\dd t\,V_t(t,r)\right|_{r=0}^{r=\infty}
	+\int_0^\infty\dd r\,\left.V_r(t,r)\right|_{t=-\infty}^{t=\infty} \,.
	\label{app:Q:density}
\end{eqnarray}
The second term vanishes because $a_r - \partial_r \varphi = 0$ as $t \to \pm \infty$ (see Eqs.~\eqref{app:lo:sigmasoft} and \eqref{app:lo:thetaposition} outside the core and $a_r, \partial_r\varphi \propto t^{-1}$ inside the core).
At the origin $r=0$, $K(0)=1$ gives $V_t(t,0)=0$, leaving only the boundary term at $r\to\infty$,
\begin{eqnarray}
	2\pi Q &=& -\lim_{r\to\infty}\int_{\mathbb R}\dd t\,V_t(t,r)
	=-\lim_{r\to\infty}\int_{\mathbb R}\dd t\,\bigl(a_t(t,r)-\partial_t\varphi(t,r)\bigr)
	=\int_{\mathbb H}\dd t\,\dd r\,f(t,r) \,,
\end{eqnarray}
where the second equality follows from $K(r\to \infty) = 0$ and the last from Stokes' theorem.
Therefore we have found that the topological charge is given by the electric flux through the half-plane.

To evaluate this flux, we use the photon equation of motion~\eqref{app:lo:sigmaeom}
\begin{eqnarray}
	&&\frac{N_f}{4\pi}\,f = 
    -\epsilon^{\alpha\beta}\partial_\alpha j_\beta
	+\frac{4\pi}{g^2}\Bqty{
    \partial^2\pqty{r^2f}
	-2\,\epsilon^{\alpha\beta}\partial_\alpha
	\bqty{K^2(a_\beta-\partial_\beta\varphi)}}\,,
	\label{app:Q:curl}
\end{eqnarray}
where $j_\beta$ is the photon current induced by the fermion zero-mode insertions and their Wilson lines, as defined in Eq.~\eqref{app:lo:insertionforces}. 
The left-hand side is the Schwinger mass term, $(4\pi/g^2)\,r^2m_{\rm Sch}^2=N_f/(4\pi)$.
Both $K^2(a_\beta-\partial_\beta\varphi)$ and $\partial_\beta(r^2f)$ vanish at the $r \to \infty$ boundary and are set to zero by the boundary conditions~\eqref{app:lo:innerbc} at $r=0$.
The terms $\epsilon^{\alpha\beta}\partial_\alpha\bqty{K^2(a_\beta-\partial_\beta\varphi)}$ and $\partial^2(r^2f)$ in Eqs.~\eqref{eq:app:Q4D} and \eqref{app:Q:curl} therefore integrate to zero over $\mathbb H$.

Consider a correlator whose insertions saturate the 't~Hooft vertex in Appendix~\ref{app:lo:integration}.
Equation~\eqref{app:lo:fermionintegral} requires $n_0$ insertions of $\psi_{\pm,f}$ and $\bar n_0$ insertions of $\bar\psi_{\pm,f}$ for each of the $N_f$ left-handed Weyl doublets.
For the Rubakov--Callan correlator, $N_f=4$\,, $n_0=1$ and $\bar n_0=0$\,.
Each $\psi$ ($\bar{\psi}$) zero mode contributes a factor $e^{-\sigma(x_i)/2}$ ($e^{\sigma(x_i)/2}$) to the insertion factor. 
Varying these factors with respect to $\sigma$ in Eqs.~\eqref{app:lo:insertionforces} and \eqref{app:lo:scalarsource} gives
\begin{eqnarray}
	\epsilon^{\alpha\beta}\partial_\alpha
	\bigl(j_\beta-j_{{\rm WL},\beta}\bigr)
	=-\frac12\sum_{i=1}^{N_f n_0}\delta^{(2)}(x-x_i)
	+\frac12\sum_{i=N_f n_0+1}^{N_f(n_0+\bar n_0)}\delta^{(2)}(x-x_i)\,,
\end{eqnarray}
where $x_i$ are the insertion points of the fermions, with the unbarred insertions listed first. 
(The curl of the Wilson-line current integrates to zero because the paths lie in the interior of $\mathbb H$.) 
Therefore we find
\begin{eqnarray}
	&&Q = \frac{1}{2\pi} \int_{\mathbb H}\dd^2x\,f = -\frac{2}{N_f} \int_{\mathbb H}\dd^2x \, \epsilon_{\alpha \beta} \partial^\alpha j^\beta = \frac{2}{N_f}\,n_A \,,
	\label{app:Q:result}
\end{eqnarray}
where $n_A$ is the total axial charge of the insertions, with each $\psi$ ($\bar\psi$) contributing $1/2$ ($-1/2$).

For the fermion currents, the identity $\gamma^\alpha\gamma_5=-i\epsilon^{\alpha\beta}\gamma_\beta$ in the conventions of Eq.~\eqref{app:lo:reducedweyl} relates the curl of the electric current to the divergence of the axial current.
We could have anticipated the relation $Q=2n_A/N_f$ from either the two-dimensional chiral anomaly or the four-dimensional axial anomaly for the $N_f$ Weyl doublets. 
When the insertions saturate the 't Hooft vertex, $2n_A=N_f(n_0-\bar n_0)$, so $Q=n_0-\bar n_0$ is an integer.
In Appendix~\ref{app: quant-charge-zm} we show that the $Q=1$ saddle has exactly $n_0=1$ and $\bar{n}_0=0$ logarithmically normalizable zero modes, precisely the modes saturated by the four insertions in the Rubakov--Callan correlator.

%%%%%%%%%%%
\subsection{Counting the zero modes}
\label{app: quant-charge-zm}
%%%%%%%%%%%

In this appendix we show a complementary argument relating the number of log-normalizable zero-modes to the instanton number $Q$.
Write $\psi_\pm^{(0)}=e^{-\sigma/2}\chi^{(0)}_\pm$, with $z=r+it$ and its conjugate $\bar z=r-it$. Then the reduced Weyl equations become 
\begin{eqnarray}
	&&2\partial_z\chi_+=P_+\chi_-\,,\qquad
	2\partial_{\bar z}\chi_-=P_-\chi_+\,,\qquad
	P_\pm\equiv \frac{K}{r}\,e^{\pm i\varphi}=\frac{m_W \,e^{\pm i\varphi}}{\sinh(m_Wr)}\,.
	\label{app:zm:dirac}
\end{eqnarray}
We begin by considering the case $\varphi=0$. Consider the total electric flux $f$ through a semi-disk $D_{\xi}^+$ with radius $\xi$ in the half plane $\mathbb H$. Through Stokes' theorem, there are two contributions:
\begin{equation}
    \int_{D_{\xi}^+} f=\int_{-\xi}^{\xi} dt \, a_t(t,0)+\int_{\partial D_{\xi}^+} a_{\alpha} dx^{\alpha}~,
\end{equation}
where $\partial D_{\xi}^+$ denotes the boundary semi-circle. The first term vanishes since $a_t(t,0)=0$ along the segment. The second term is the contour integral of the flux enclosed by the semi-circle with radius $\xi$, since $a_t=-\partial_r\sigma$\,, $a_r=\partial_t\sigma$\,. The semicircle integral then yields 
\begin{equation}
\label{logfix}
\int_{D_{\xi}^+} f = \int_{-\pi/2}^{\pi/2} \xi\partial_\xi\sigma\, d\theta \;\xrightarrow{\ \xi\to\infty\ }\; 2\pi Q~,
\end{equation}
where $\theta=\rm{arg}(z)$.  Therefore, we have
\begin{eqnarray}
	\sigma\to2Q\ln(m_W \xi)~,\qquad e^{-\sigma/2}\sim(m_W\xi)^{-Q}~,\qquad \xi\to\infty\,.
	\label{app:zm:photon}
\end{eqnarray}

We now consider the contribution from $\chi$ to the counting. For $\varphi=0$, we have $P_\pm=p=K(r)/r$ and
\begin{eqnarray}
	&&4\,\partial_z\partial_{\bar z}\ln(p/m_W)=\left(\ln(p/m_W)\right)''=p^2\,.
	\label{app:zm:liouville}
\end{eqnarray}
Then the two components decouple: eliminating $\chi_-$ or
$\chi_+$ from Eq.~\eqref{app:zm:dirac} gives
\begin{eqnarray}
	&&\partial_z\Bigl[p\,\partial_{\bar z}\frac{\chi_+}{p}\Bigr]=0\,,\qquad
	\partial_{\bar z}\Bigl[p\,\partial_z\frac{\chi_-}{p}\Bigr]=0\,.
	\label{app:zm:decoupled}
\end{eqnarray}
These are solved by $\chi_+=pB(z)/2$ and by $\chi_-=pA(\bar z)/2$, with $B$
holomorphic and $A$ antiholomorphic on $\mathbb H$. Primes on $B$ and $A$
denote derivatives with respect to their arguments.
Equation~\eqref{app:zm:dirac} supplies the other component in each case:
\begin{eqnarray}
	&&\chi_+=\frac{p}{2}\,B+\frac{1}p\,\partial_{\bar z}(pA)\,,\qquad
	\chi_-=\frac{1}p\,\partial_z(pB)+\frac{p}{2}\,A\,.
	\label{app:zm:general}
\end{eqnarray}
Note that this is a general solution since $A,B$ are arbitrary.

Near the origin $p=1/r+O(m_W^2r)$, such that $\chi_\pm=\pm(B-A)/2r+O(1)$. The
solution is regular if and only if $A=B$ on the line $\bar z=-z$, that is, if
and only if $B$ continues to an entire function and $A(\bar z)=B(-\bar z)$:
\begin{eqnarray}
	&&\chi_-=B'(z)-\frac{m_W[\cosh(m_Wr)B(z)-B(-\bar z)]}{2\sinh(m_Wr)}\,,\qquad
	\chi_+(t,r)=-\chi_-(t,-r)\,.
	\label{app:zm:regular}
\end{eqnarray}
As long as $\chi$ grows no
faster than a power of $|z|$, the entire function \(B''-m_W^2B/4\) has polynomial growth and is therefore a polynomial by Liouville's theorem. Since $B=e^{\pm m_W z/2}$ drop
out of Eq.~\eqref{app:zm:regular}, $B$ itself can be taken to be a polynomial, and $\chi_-\to B'(z)-m_WB(z)/2$ for $m_Wr\gg1$. With Eq.~\eqref{app:zm:photon}, this fixes the zero mode counting as 
\begin{eqnarray}
	&&|\psi_\pm|\sim(m_W\xi)^{\,n-Q}\,,\qquad n\equiv\deg B=0,1,2,\dots\,,
	\label{app:zm:tower}
\end{eqnarray}
up to an overall mode normalization. Here normalizability refers to the
physical spinor norm
\begin{eqnarray}\label{eq:app:spinornorm}
    \|\psi\|^2=\int_{\mathbb H}\dd{t}\,\dd{r}\,(|\psi_+|^2+|\psi_-|^2) \, .
\end{eqnarray}

For general phase, we can use a homotopy argument by studying a deformation of the BPS Dirac operator. Consider a continuous family of operators:
\begin{eqnarray}
	\slashed{D}_\lambda=\slashed{D}_{\rm BPS}+V_\lambda\,,\qquad
	V_\lambda=\frac Kr\pqty{e^{-i\lambda\gamma_5\varphi}-1}\,,
	\qquad0\le\lambda\le1\,.
	\label{app:zm:family}
\end{eqnarray}
Fix the asymptotic amplitude $c$ and write
$\chi_\lambda=c\chi_0+\delta\chi_\lambda$. The decaying correction then obeys
\begin{equation}
    \delta\chi_\lambda
    =-G_{\rm BPS}'V_\lambda(c\chi_0+\delta\chi_\lambda),
    \label{app:zm:phasecontinuation}
\end{equation}
where $G_{\rm BPS}'$ is the non-zero mode fermion propagator in the BPS background~\eqref{app:lo:loopscattering}. 
Each iteration of Eq.~\eqref{app:zm:phasecontinuation} therefore brings a factor of at most $\pi^2\|V_\lambda/(2m_W K(r))\|_\infty$.
As we work in the limit $g \ll 1$, this factor is less than one, and the iterations converge to a unique correction for each asymptotic amplitude $c$. 
It determines a unique solution deformed from $\chi_0$. 
When dressed with $e^{-\sigma/2}$, the norm~\eqref{eq:app:spinornorm} of $\psi$ remains log-normalizable, while for the conjugate solution, the dressing is $e^{\sigma/2}$ and there are no log-normalizable zero modes. 
All told, per Weyl flavor,
\begin{eqnarray}
	n_0=1\,,\qquad\bar n_0=0\,,\qquad Q=1\,,\quad0\le\lambda\le1\,.
	\label{app:zm:count}
\end{eqnarray}
Therefore we have shown that for the saddle configuration \eqref{app:lo:sigmasoft} and \eqref{app:lo:thetaposition} of the Rubakov--Callan correlator, the number of log-normalizable zero modes is precisely determined by the sourced instanton number.

%%%%%%%%%%%
\section{LSZ reduction}
\label{app:radial-lsz}
%%%%%%%%%%%
%
We apply a two-dimensional Lehmann--Symanzik--Zimmermann (LSZ) reduction to the radial correlator $C^{2D}$ in Eq.~\eqref{eq:correlator solution} to find the scattering amplitudes of the lowest-partial-wave fermions $\psi_f$ in the Rubakov--Callan process.
We continue from Euclidean to Lorentzian time, $t_k\to i t_k^{\rm L}$, and amputate the external propagators to obtain
\begin{eqnarray}
	\delta(E_1+E_2-E_3-E_4)\,{\cal M}_{\rm RC}
	&=&-\frac{1}{(2\pi)^2}\lim_{r_k\to\infty}
	\int\!\dd^4t^{\rm L}\,
	e^{-i\sum_{i=1}^2E_i(t_i^{\rm L}+r_i)+i\sum_{j=3}^4E_j(t_j^{\rm L}-r_j)}C^{2D}\,,
	\label{eq:cr-lsz-projection}
\end{eqnarray}
where $\dd^4t^{\rm L}\equiv\prod_{k=1}^4\dd{t_k^{\rm L}}$.
The quantity ${\cal M}_{\rm RC}$ is the Rubakov--Callan scattering amplitude in the lowest partial wave, with incoming energies $E_{1,2}$ and outgoing energies $E_{3,4}$.
The radial states are normalized by $\braket{E'}{E}=\delta(E'-E)$, with the overall phase chosen so that the low-energy amplitude is real and positive.

In the low-energy limit $E_i\ll g^2m_W/(4\pi^2)$, the result is independent of $g$ and $m_W$, as expected from the Abelian-instanton description, which is insensitive to the core dynamics in this limit~\cite{Csaki:2024ajo}:
\begin{eqnarray}
	{\cal M}_{\rm RC}
	&=&\frac{2}{\pi^2\sqrt{\max(E_1E_2,E_3E_4)}}\,
	\mathcal K\pqty{\frac{\min(E_1E_2,E_3E_4)}{\max(E_1E_2,E_3E_4)}}\,.
	\label{eq:cr-far-amplitude}
\end{eqnarray}
Here $\mathcal K(x)=\int_0^{\pi/2}\dd\varphi/\sqrt{1-x\sin^2\varphi}$ is the complete elliptic integral and should not be confused with the monopole profile $K(r)$.

For $E_i,\abs{E_1-E_3},\abs{E_2-E_3}\gg g^2m_W/(4\pi^2)$, each large-frequency factor in the Fourier transform falls as $\omega^{-2}$.
The leading contributions contain three such factors, while the fourth is integrated over frequencies of order $g^2m_W$.
This gives
\begin{eqnarray}
	{\cal M}_{\rm RC}
	&\simeq&\frac{i e^{2\gamma_E}}{8}\pqty{\frac{g^2m_W}{4\pi^2}}^5
	\bqty{\frac{1}{(E_2-E_3)^2\min(E_1E_3,E_2E_4)^2}
	+\frac{1}{(E_1-E_3)^2\min(E_1E_4,E_2E_3)^2}}\,,
	\label{eq:cr-rotor-amplitude}
\end{eqnarray}
where $\gamma_E$ is the Euler--Mascheroni constant.
The points $E_1=E_3$ and $E_2=E_3$ require care and can be treated, for example, by smearing in energy.
Let $E$ be a common physical energy scale, with all ratios $E_i/E$ fixed.
For the transform of Eq.~\eqref{eq:correlator solution}, away from these points, the two limits give
\begin{eqnarray}
	{\cal M}_{\rm RC}&\propto&
	\begin{cases}
		E^{-1}\,, & E\ll g^2m_W\,,\\[3pt]
		\pqty{g^2m_W/E}^{5}E^{-1}\,, & 
        g^2m_W\ll E\,.
	\end{cases}
	\label{eq:cr-scaling}
\end{eqnarray}

Therefore, for $E\gg m_W$ at weak coupling, the amplitude obtained from Eq.~\eqref{eq:correlator solution} falls off to zero roughly as  $E^{-6}$ at fixed generic energy ratios.

\bibliography{bibliography}

\end{document}